\documentclass[superscriptaddress,twocolumn]{revtex4}
\usepackage{amsmath,lscape,epsfig}

\def\ii{\'{\i}}
\def\beq{\begin{equation}}
\def\eeq{\end{equation}}
\def\beqa{\begin{eqnarray}}
\def\eeqa{\end{eqnarray}}
\def\ban{\begin{eqnarray*}}
\def\ean{\end{eqnarray*}}
\def\bi{\begin{itemize}}
\def\ei{\end{itemize}}

\begin{document}

\title{Low density expansion and isospin dependence of nuclear energy functional: comparison between relativistic and Skyrme models}

\author{C. Provid\^encia}
\affiliation{Centro de F\ii sica Te\'orica - Dep. de F\ii sica -
Universidade de Coimbra - P-3004 - 516 - Coimbra - Portugal}
\author{D.P.Menezes}
\affiliation{Depto de F\'{\i}sica - CFM - Universidade Federal de
Santa Catarina  Florian\'opolis - SC - CP. 476 - CEP 88.040 - 900
- Brazil}
\author{L. Brito}
\affiliation{Centro de F\ii sica Te\'orica - Dep. de F\ii sica -
Universidade de Coimbra - P-3004 - 516 - Coimbra - Portugal}
\author{Ph. Chomaz}
\affiliation{GANIL (DSM-CEA/IN2P3-CNRS), B.P. 5027, F-14076 Caen
C\'edex 5, France}

\begin{abstract}
In the present work we take the non relativistic limit of
relativistic
 models and compare the obtained functionals with the usual Skyrme
parametrization. Relativistic models with both constant couplings
and with density dependent couplings are considered. 
While some
models present very good results already at the lowest order in
the density, models with non-linear terms only reproduce the
energy functional if higher order terms are taken into account in
the expansion.
\end{abstract}

\maketitle

\vspace{0.0cm} PACS number(s):
{21.65.+f,24.10.Jv,21.30.-x,21.60.-n}
\vspace{0.50cm}

\section{Introduction}

Both non-relativistic and relativistic phenomenological nuclear
models are nowadays used to describe successfully nuclear matter
and finite nuclei  within a density functional formalism. One of
the most commonly used non-relativistic models, the Skyrme force
model, has been extensively used in the literature since the work
of Vautherin and Brink \cite{brink}.
The Skyrme effective
interaction has been improved since its first version
\cite{original} in order to account for the different properties
of nuclei, not only along the stability line, but also for exotic
nuclei from the proton to the neutron drip line and 
to yield a good description of neutron stars \cite{chabanat}.
Along this line, many different versions of the Skyrme effective
force have been developed and tested and the importance of the
asymmetry discussed \cite{chabanat,sple}. In \cite{akmal}, the
results of the variational microscopic calculations, in which
many-body and some relativistic corrections were included,  were
parametrized. In \cite{sple} the Skyrme model parameters were
chosen so as to fit the EoS given in \cite{akmal}. In \cite{stone}
87 different parametrizations, which give similar results for
finite nuclei experimental observables, were checked against
neutron star properties and 60 of them were ruled out. Some of the
non-relativistic models include also a three-body force
\cite{3body} in order to solve the causality problem, reproduce
saturation properties of nuclear matter and improve the
description of the symmetry energy.

Relativistic models, on the other hand, are advantageous if high
density matter, such as the one existing in compact stars, is
described, and in particular no lack of causality arises as
density increases. As seen in \cite{stone}, neutron star
properties depend on the correct choice of the equation of state
(EoS) and the same problem appears within relativistic models,
which can be parametrized in different ways, all of them giving
similar results for finite nuclei and nuclear matter or neutron matter.
Once the EoS
is extrapolated to high densities, quantities as symmetry energy,
for example, depend a lot on the parameter set chosen \cite{alex,
compact1}. At subsaturation densities, there is an unstable
region, which varies a lot and present different behaviors
according to the parametrization chosen
\cite{inst,spinodal,modos}. Some relativistic models introduce the
density dependence through the couplings of baryons to mesons
\cite{TW,gaitanos,br} and once again the instability region at low
densities and the EoS at high densities are sensible to the model
considered \cite{spinodal,br}. We point out that the saturation mechanism of 
relativistic
  and non-relativistic models is different. For the first ones saturation is attained due to
  the relativistic quenching of the scalar field. On the other hand, non-relativistic models
  have to introduce three-body repulsive interactions in order to describe saturation correctly.

At very low densities both, the relativistic and the
non-relativistic approaches predict 
a liquid-gas phase transition region for nuclear matter leading, 
for dense star matter to a non-homogeneous phase
commonly named {\em pasta phase}, formed by a competition between
the long-range Coulomb repulsion and the short-range nuclear
attraction \cite{pasta}. 

In the past some attempts have already been made in order to
compare nuclear matter and finite nuclei properties obtained both
with relativistic and non-relativistic models
\cite{bao-li,spinodal,ring97} but there is no clear or obvious
explanations for the differences. In the following we compare the
Skyrme effective force with relativistic nuclear mean-field models
at subsaturation densities with the goal to directly compare the energy 
functional. We expect that the same physics should be contained in both 
approaches at these
  densities and therefore it maybe fruitful to compare both types of models in this range of densities.

 We start with a brief review of the
Skyrme parametrizations of the nuclear energy density functional.
Next we consider
relativistic mean-field  models with constant couplings, namely
the initially proposed  parametrizations including only linear
terms for the meson contributions \cite{qhd1} and parametrizations
with non-linear terms \cite{nl3,tm1,liu}. Then we extend our
investigation to density dependent models
\cite{TW,gaitanos,inst,hor,bunta}.  We compare the
non-relativistic limit of their binding energies with the binding
energy functional obtained with the non-relativistic Skyrme model.
Various levels of the approximation are then discussed.

%PC \subsection{\protect\smallskip Skyrme functional}
\section{\protect\smallskip Skyrme functional}

 The non relativistic Skyrme energy functional is defined by
\[
B_{\rm
Skyrme}=\mathcal{K}+\mathcal{H}_{0}+\mathcal{H}_{3}+\mathcal{H}_{\rm
eff},
\]
where $\mathcal{K}$ is the kinetic-energy density, 
$\mathcal{H}%
_{0}$ a density-independent two-body term, $\mathcal{H}_{3}$ a
density-dependent term, and $\mathcal{H}_{\rm eff}$ a
momentum-dependent term:
\begin{eqnarray}
\mathcal{K} &=&\frac{\tau }{2M}, \\
\mathcal{H}_{0} &=&C_{0}\rho ^{2}+D_{0}\rho _{3}^{2}, \\
\mathcal{H}_{3} &=&C_{3}\rho ^{\sigma +2}+D_{3}\rho ^{\sigma }\rho
_{3}^{2},
\\
\mathcal{H}_{\rm eff} &=&C_{\rm eff}\rho \tau +D_{\rm eff}\rho
_{3}\tau _{3}, \label{skyrme1}
\end{eqnarray}
where $\rho =\rho _{p}+\rho _{n}$ is the total baryonic density,
$\rho _{p}$ and $\rho _{n}$ being the proton and the neutron densities 
respectively, 
$\rho_3 =\rho _{p}-\rho_{n}$, the isovector density,
$\tau=\sum_{i=p,n} \tau_i$, the total kinetic density, 
$\tau_i=k_{F_i}^5/5 \pi^2$
being the kinetic density of each type of particles $i$ with a Fermi 
momentum $k_{F_i}$, 
and  $\tau_3=\tau_p-\tau_n$  the isovector kinetic term. 
We introduce the following quantities:
$B_1$ the symmetric matter potential energy, 
$B_3$ the potential part of the symmetry energy and
$M_i^*$ the proton $(i=p)$ or neutron $(i=n)$ effective masses,
\begin{eqnarray}
B_{1}(\rho ) &=&C_{0}{\rho }^{2}+C_{3}\rho ^{\sigma +2},  \label{B1S} \\
B_{3}(\rho ) &=&D_{0}+D_{3}\rho ^{\sigma },  \label{B3S} \\
M_{i}^{*^{-1}} &=&M^{-1}+2 C_{\rm eff}\rho +\tau _{3i}~2 D_{\rm
eff}\rho _{3}, \label{masseffS}
\end{eqnarray}
such that
\begin{equation}
B_{\rm Skyrme}=\sum_{i=p,n}{\frac{\tau
_{i}}{2{M_{i}^{*}}}+}B_{1}(\rho )+\rho _{3}^{2}B_{3}(\rho )
\label{skyr}.
\end{equation}

The coefficients $C_{i}$ and $D_{i}$, associated respectively with
the symmetry and asymmetry contributions, are linear combinations
of the traditional Skyrme parameters :
\[
\begin{array}{ll}
C_{0} & =\ \ 3t_{0}/8 \\
D_{0} & =-t_{0}(2x_{0}+1)/8\\
C_{3} & =\ \ t_{3}/16 \\
D_{3} & =-t_{3}(2x_{3}+1)/48 \\
C_{\rm eff} & =\ \ [3t_{1}+t_{2}(4x_{2}+5)]/16 \\
D_{\rm eff} & =\ \ [t_{2}(2x_{2}+1)-t_{1}(2x_{1}+1)]/16
\end{array}
\]
and $\sigma$ parametrizes the density dependent term. The $\sigma$
 exponent has a direct effect on the incompressibility. A decrease in
 $\sigma$ generates a lower value for the incompressibility \cite{chabanat}.
Using $\rho _{i}=k_{F_{i}}^{3}/3\pi ^{2}$ the kinetic terms can
be directly related to the respective density 
\[
\tau _{i}=a(\rho_{i})^{5/3}
\]
where we have introduced $a=3^{5/3}\pi ^{4/3}/5$. Looking at small
asymmetries $\rho _{_{_{n}^{p}}}=\left( \rho \pm \rho _{3}\right) /2$ we get

\begin{eqnarray}
\tau  &=&a\left( \rho ^{5/3}+\frac{5}{9}\rho ^{-1/3}\rho _{3}^{2}\right)  \\
\tau _{3} &=&\frac{5a}{3}\rho ^{2/3}\rho _{3}
\end{eqnarray}
Then it is easy to decompose the kinetic energy density $\mathcal{K}_{\rm eff}=%
\mathcal{K}+\mathcal{H}_{\rm eff}$ into a symmetric matter component and a
symmetry kinetic energy 
\[
\mathcal{K}_{\rm eff}=\mathcal{K}_{\rm eff_{1}}+\rho _{3}^{2}\mathcal{K}_{\rm eff_{3}}
\]
leading to 
\begin{eqnarray*}
\mathcal{K}_{\rm eff_{1}} &=&a\rho ^{5/3}(\frac{1}{2M}+C_{\rm eff}\rho ) \\
\mathcal{K}_{\rm eff_{3}} &=&a\frac{5}{9}\rho ^{-1/3}\left[\frac{1}{2M}%
+(C_{\rm eff}+3D_{\rm eff})\rho \right]
\end{eqnarray*}
The isoscalar part of the effective mass $C_{\rm eff}$ directly contributes to $%
\mathcal{K}_{\rm eff_{1}}$ but also contributes to the symmetry energy because of
the asymmetry in the Fermi energy. Indeed as in the Fermi gas model the two
first terms of $\mathcal{K}_{\rm eff_{3}}$ can be recasted as $\mathcal{K}%
_{\rm eff_{3}}=5\mathcal{K}_{\rm eff_{1}}/9\rho ^{2}.$ The last term is an
additional term coming from the isovector part of the effective mass. 
These various contributions to the kinetic energy play an important role
both in the binding energy and saturation properties of symmetric matter  
and in the isospin dependence usually discussed in terms of symmetry energy.
When comparing with relativistic approaches the discussion of the kinetic 
part becomes even more important since the role played by the effective 
mass and interaction are known to be in general rather different.  
In Table \ref{tab:properties} we present the nuclear matter saturation 
properties obtained 
with the Skyrme forces used in the present work.
We consider a conventional Skyrme interaction SIII, one of the recent Skyrme-Lyon interactions
designed to describe neutron-rich matter SLy230a \cite{chabanat},
the  NRAPR parametrization which stands for the Skyrme
interaction parameters obtained from a fitting to the EoS of a
microscopic model \cite{akmal,sple} and the LNS parametrization which refers to a recently
Skyrme-like parametrization proposed in the framework of the
Brueckner-Hartree-Fock approximation for nuclear matter
\cite{lns}.

\begin{table}[c]
  \centering
  \begin{tabular}{cccccc}
\hline
    Model & $B/A$ & $\rho_0$ & $K$ & ${\cal E}_{sym}$ & $M^*/M$\\
 & (MeV) & (fm$^{-3}$)& (MeV)& (MeV)&\\
\hline
SIII \cite{chabanat}& 15.851 & 0.145 & 355.5 & 28.16 & 0.76 \\
SLy230a \cite{chabanat} & 15.989 & 0.16 & 229.87 & 31.97 & 0.697 \\
NRAPR \cite{sple}& 15.86& 0.16&225.7&32.79&0.7\\
LNS \cite{lns}&15.32&0.175&210.85&33.4&0.825\\
\hline
  \end{tabular}
\caption{Nuclear matter properties of the Skyrme forces used in the present 
work }
  \label{tab:properties}
\end{table}

\section{\protect\smallskip Relativistic approaches}

\subsection{\protect\smallskip The lagrangian}

In this section we consider four mean-field relativistic models,
which we denote by QHD-II \cite{qhd1}, NL3 \cite{nl3}, TM1
\cite{tm1} and NL$\delta $ \cite{liu}, with constant coupling
parameters described by the Lagrangian density of the linear and
non-linear Walecka models (NLWM), with the possible inclusion of
the $\delta $ mesons, given by: \smallskip
\begin{equation}
\mathcal{L}_{NLWM}=\sum_{i=p,n}\mathcal{L}_{i}\mathcal{\,+L}_{{\sigma }}%
\mathcal{+L}_{{\omega }}\mathcal{+L}_{{\rho
}}\mathcal{+L}_{{\delta }}, \label{lagdelta}
\end{equation}
where the nucleon Lagrangian reads

\begin{equation}
\mathcal{L}_{i}=\bar{\psi}_{i}\left[ \gamma _{\mu }iD^{\mu
}-\mathcal{M}^{*}\right] \psi _{i},  \label{lagnucl}
\end{equation}
with 
\begin{eqnarray}
iD^{\mu } &=&i\partial ^{\mu }-g_{v}V^{\mu }-\frac{g_{\rho }}{2}{\vec{\tau}}%
\cdot \vec{b}^{\mu },  \label{Dmu} \\
\mathcal{M}^{*} &=&M-g_{s}\phi -g_{\delta }{\vec{\tau}}\cdot
\vec{\delta}. \label{Mstar}
\end{eqnarray}
The isoscalar part is associated with the scalar sigma ($\sigma $) field, $%
\phi $, and the vector omega ($\omega $) field, $V_{\mu }$, while
the isospin dependence comes from the isovector-scalar delta
($\delta $) field, $\delta ^{i}$, and the isovector-vector rho
($\rho $) field, $b_{\mu }^{i}$ (where $\mu $ is the 4 dimensional
space-time indices and $i$ the 3D isospin direction indices).  The
associated Lagrangians are
\begin{eqnarray*}
\mathcal{L}_{{\sigma }} &=&+\frac{1}{2}\left( \partial _{\mu }\phi \partial %
^{\mu }\phi -m_{s}^{2}\phi ^{2}\right)-\frac{1}{3!}\kappa \phi ^{3}-\frac{1}{4!}%
\lambda \phi ^{4},  \\
\mathcal{L}_{{\omega }} &=&-\frac{1}{4}\Omega _{\mu \nu }\Omega ^{\mu \nu }+%
\frac{1}{2}m_{v}^{2}V_{\mu }V^{\mu }+\frac{1}{4!}\xi
g_{v}^{4}(V_{\mu
}V^{\mu })^{2}, \\
\mathcal{L}_{{\delta }} &=&+\frac{1}{2}(\partial _{\mu }\vec{\delta}\partial %
^{\mu }\vec{\delta}-m_{\delta }^{2}{\vec{\delta}}^{2}\,), \\
\mathcal{L}_{{\rho }} &=&-\frac{1}{4}\vec{B}_{\mu \nu }\cdot
\vec{B}^{\mu \nu }+\frac{1}{2}m_{\rho }^{2}\vec{b}_{\mu }\cdot
\vec{b}^{\mu },
\end{eqnarray*}
where $\Omega _{\mu \nu }=\partial _{\mu }V_{\nu }-\partial _{\nu
}V_{\mu }$
, $\vec{B}_{\mu \nu }=\partial _{\mu }\vec{b}_{\nu }-\partial _{\nu }\vec{b}%
_{\mu }-g_{\rho }(\vec{b}_{\mu }\times \vec{b}_{\nu })$, and $g_{j}$ and $%
m_{j}$ are respectively the coupling constants of the mesons
$j=s,v,\rho ,\delta $ with the nucleons and their masses.
Self-interacting terms for the $\sigma $-meson are also included
in the three parametrizations, $\kappa $
and $\lambda $ denoting the corresponding coupling constants. The $\omega $%
-meson self-interacting term, with the $\xi $ coupling constant,
is present in the TM1 parametrization. In the above Lagrangian
density $\vec{\tau}$ is the isospin operator. The terms involving
the $\delta $ meson are only present in the NL-$\delta$ model and
non-linear terms are not present in the simplest version of the
model QHD-II.
In Table \ref{tab:propertiesrel} we present the nuclear matter saturation 
properties obtained with the relativistic models used in the present work.

\begin{table}[c]
  \centering
  \begin{tabular}{cccccc}
\hline
    Model & $B/A$ & $\rho_0$ & $K$ & ${\cal E}_{sym}$ & $M^*/M$\\
 & (MeV) & (fm$^{-3}$)& (MeV)& (MeV)&\\
\hline
Walecka \cite{qhd1}& 15.75&0.192&540&22.1&0.556\\
NL3 \cite{nl3} &16.3& 0.148&272 &37.4&0.60\\
TM1 \cite{tm1}&16.3&0.145& 281& 36.9& 0.63\\
NL$\delta$ \cite{liu} &16.0&0.160&240 & 30.5& 0.60\\
TW \cite{TW}&16.3 & 0.153&240& 32.0& 0.56 \\
 DDH$\delta$ \cite{gaitanos} &16.3&0.153& 240& 25.1& 0.56\\
\hline
  \end{tabular}
  \caption{Nuclear matter properties of the relativistic models used in the 
present work }
 \label{tab:propertiesrel}
\end{table}

\subsection{Equilibrium}

At equilibrium the fermion distribution is a Fermi-Dirac
distribution function for nucleons (+) an antinucleons (-)

\begin{equation}
f_{i\pm }={1}/\{1+\exp ((\epsilon_{i}^{*}(\mathbf{p})\mp \nu
_{i})/T)\}\;, \label{distf}
\end{equation}
where $\epsilon_{i}^{*}=\sqrt{\mathbf{p}^{2}+{M_{i}^{*}}^{2}}$ with the effective mass
\begin{equation}
M_{i}^{*}=M-g_{s}~\phi _{0}-\tau _{3i}~g_{\delta }~\delta _{3},
\label{meff}
\end{equation}
In the equilibrium, the effective chemical potentials are 
defined by
\begin{equation}
\nu _{i}=\mu _{i}-g_{v}V_{0}-\frac{g_{\rho }}{2}~\tau _{3i}~b_{0},
\label{chemical}
\end{equation}
where $\tau _{3i}=\pm 1$ is the isospin projection for the protons and
neutrons respectively. At zero temperature the distribution
function reduces to a simple step function
\begin{equation}
f_{i}=\theta (p_{_{F_{i}}}^2- p^2),  \label{fT0}
\end{equation}
with 
\begin{equation}
p_{_{F_{i}}}=\sqrt{\nu _{i}^{2}-{M_{i}^{*}}^{2}}.\label{pfermi}
\end{equation}

\smallskip Let us introduce $\phi _{0}$, $V_{0}$, $\delta _{3}$ and $b_{0}$
the values of the scalar ($\sigma $), the vector ($\omega $),
isovector scalar ($\delta $) and the isovector vector  ($\rho $)
fields, obtained from the meson equations of motion, considered as
static and uniform classical fields
\begin{eqnarray}
m_{s}^{2}\phi _{0}+\frac{1}{2}\kappa \phi
_{0}^{2}+\frac{1}{6}\lambda \phi
^{3} &=&g_{s} \rho_s,   \label{Eq:sigma} \\
m_{v}^{2}V_{0} + \frac {1}{6} \xi g_v^4 V_0^3 &=&g_{v} \rho,
\label{EQ:omega} \\
m_{\delta }^{2}\delta_3  &=&g_{\delta} \rho_{s3},   \label{EQ:delta} \\
m_{\rho }^{2}b_{0} &=&\frac{g_{\rho }}{2} \rho_3, \label{EQ:rho}
\end{eqnarray}
where the baryonic and scalar densities read:
\begin{equation}
\rho _{i}=2\int \frac{d^{3}\mathbf{p}}{\left( 2\pi \right)
^{3}}\theta (p_{_{F_{i}}}^2-p^2),  \label{rhoi}
\end{equation}

\begin{equation}
\rho _{s_{i}}=2\int \frac{d^{3}\mathbf{p}}{\left( 2\pi \right)
^{3}}\,\frac{ M_{i}^{*}}{\sqrt{p^{2}+{M_{i}^{*}}^{2}}}\theta
(p_{_{F_{i}}}^2-p^2), \label{rhoscalar}
\end{equation}
and $\rho = \rho _{p}+\rho_{n}\,$, $\rho_{3} = \rho
_{p}-\rho_{n}\, $, $\rho_{s} = \rho _{sp}+\rho_{sn}\, $,
$\rho_{s3} = \rho _{sp}-\rho_{sn} $.
The above ensemble of equations 
for the effective mass (\ref{meff}), the fields 
(\ref{Eq:sigma})-(\ref{EQ:rho}) and the densities
(\ref{rhoi})-(\ref{rhoscalar}) related to the Fermi momentum (\ref{pfermi}) 
defines entirely the state  in a self consistent
way.  
In the sequel we introduce the constants $c_{\rho }=g_{\rho
}^{2}/8m_{\rho }^{2}$ and $c_{\alpha }=g_{\alpha
}^{2}/2m_{\alpha}^{2}$, with $\alpha=s,v,\delta$. 

The energy density is then given :
\begin{equation}
E=K_{p}+K_{n}+E_{\sigma }+E_{\omega }+E_{\delta }+E_{\rho },
\label{EQ:E}
\end{equation}
with
\begin{eqnarray}
K_{i} &=&2\int \frac{d^{3}\mathbf{p}}{\left( 2\pi \right) ^{3}}{\sqrt{p^{2}+{%
M_{i}^{*}}^{2}}}\,\theta (p_{_{F_{i}}}^2-p^2),  \label{EQ:EK} \\
E_{\sigma } &=&\frac{m_{s}^{2}}{2}\phi _{0}^{2}+\frac{1}{6}\kappa \phi_0 ^{3}+%
\frac{1}{24}\lambda \phi_0 ^{4},  \label{EQ:Esigma} \\
E_{\omega } &=&\frac{m_{v}^{2}}{2}V_{0}^{2}+ \frac{1}{8} \xi g_v^4
V_0^4 \label{EQ:Eomega}
\\
E_{\delta } &=&\frac{m_{\delta }^{2}}{2}\delta _{3}^{2}=c_{\delta
}
{\rho _{s3}}^{2},  \label{EQ:Edelta} \\
E_{\rho } &=&c_{\rho }\rho _{3}^{2}, \label{EQ:Erho}
\end{eqnarray}
Having solved the field's self-consistent equations entirely defines  
the system energy functional. 

In order to compare the relativistic approaches with the non relativistic 
one  we also look at the binding energy density defined as
\begin{equation}
B={E}-M\rho. \label{bindef}
\end{equation}
Isolating the kinetic contribution
\begin{equation}
T_i= K_i - M_{i}^* \rho,
\end{equation}
we can define an interaction part  
\begin{equation}
B_{{\rm exact}}(\rho,\rho_3)=B-\sum_i T_i ,
\label{bexact}
\end{equation}
which can be decomposed into an isoscalar part
\begin{equation}
B_{1 \rm exact}(\rho)=B_{\rm exact}(\rho,\rho_3=0) \label{b1exact}
\end{equation}
and isovector potential energy contributions
\begin{equation}
B_{3 \rm exact}(\rho,\rho_3)=\frac{B_{\rm exact}(\rho,\rho_3)-B_{1
\rm exact}(\rho)} {\rho_3^2}.\label{b3exact}
\end{equation}
Using these notations the binding energy from relativistic approaches can be 
recasted as in the Skyrme case
\begin{equation}
B=\sum_i T_i + B_{1 \rm exact}(\rho) + B_{3
\rm exact}(\rho,\rho_3)\,\rho_3^2. \label{Bexactsky}
\end{equation}
 
 We also separate  the effective mass into the isoscalar and the isovector channels
and write
\begin{equation}
M_i^*= M+M_1 +\tau_{3i} \rho_3 M_3.
\label{nlmeff}
\end{equation}

\section{\protect\smallskip Direct Comparison of energy functional}

The first idea to make a bridge between the classical and relativistic 
models is to directly compare the various terms entering in the associated 
energy functionals.
Indeed, considering the Kohn-Sham theorem,  models leading to the 
same energy functional are strictly equivalent as far as physical results 
are concerned. However, in the previous sections, we have seen that for 
spin saturated static { matter } the Skyrme energy is a functional of the 
isoscalar and isovector particle and kinetic densities. The case of the 
relativistic models is more complex since the proton and neutron baryonic
and scalar densities appear as well as the relativistic kinetic energy 
density. This makes a direct comparison of the functional impossible. 
In this paper we thus adopt two strategies: 
i) In the next section we develop a non-relativistic approximation to these 
densities in order to reduce them to the Skyrme like local densities. Then a 
direct comparison of the functionals at low densities is possible and will teach us a lot about 
the connections between those two classes of models. A low density 
expansion of the energy per particle gives
\begin{eqnarray*}
  E/\rho &=& M + \frac{3k_F^2}{10 M}
            {- \frac{3k_F^4}{56 M^3}}
             + \cdots 
      +\Big( {g_v^2\over 2m_v^2} - { g_s^2\over 2 m_s^2}\Big)\, \rho\nonumber\\
   & &   + {{ g_s^2\over m_s^2}\, {\rho\over M}\bigg[ {3k_F^2\over 10 M}
      - \cdots \bigg]}
      {+ \Big({ g_s^2\rho\over m_s^2 M}\Big)^{\mkern-2mu 2}
         \bigg[{3k_F^2\over 10 M}
          -  \cdots \bigg]}\\
   & &         { + {\cal O} \left[\Big({ g_s^2\rho\over m_s^2 M}\Big)^{\mkern-2mu 3}\right].}
\end{eqnarray*}
where the terms of the type $ \Big({ g_s^2\rho\over m_s^2
M}\Big)^{\mkern-2mu n}, \,\ n=1,2,3,...$ have origin on  the expansion
of the scalar density and are equivalent to $n$-body terms of the
Skyrme interaction generally described by fractionary exponents.
ii) Secondly, we look at the contributions of the various energy terms 
to the nuclear matter EoS thus reducing the functionals of both models to 
particle densities. With this reduction, exact results for the various models 
can be directly compared.

\begin{figure}[t]
\begin{center}
\begin{tabular}{c}
 \includegraphics[width=6.5cm,angle=0]{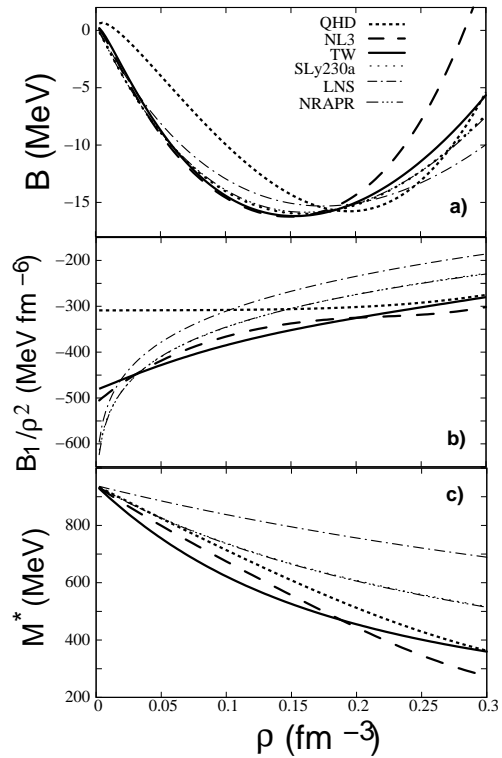} \\
\end{tabular}
\end{center}
\caption{Symmetric matter: a) Binding energy; b) Isoscalar
contribution of the interaction divided by $\rho^2$,  $B_{1\rm exact}/\rho^2$; c) Effective 
mass for several relativistic and non-relativistic models.}
\label{sym}
\end{figure}

Let us first focus on the properties of symmetric matter. In Figs.
\ref{sym}a), \ref{sym}b) and \ref{sym}c) we show respectively the 
binding energy, the isoscalar contribution of the interaction $B_{1\rm exact}$ and the effective mass for several relativistic 
and non-relativistic models.
We include the results of the Walecka (QHD-II) model just for comparison
although we know it does not reproduce well the properties of nuclear
matter at saturation. In particular, it has a very high compressibility.
A comparison of the binding energies plotted in Fig. \ref{sym}a) shows
that
i) although the saturation point is quite close in energy and densities for all models,
there is some discrepancy which is related to the way the
parametrizations were fitted;
ii) relativistic and non-relativistic models show different behaviors
above an below saturation density: at subsaturation densities
relativistic models show generally more binding (except for the
Walecka parametrization) and above saturation density the binding
energy decreases faster with density within these models. The
parametrization with density dependent coefficients (TW) has a behavior
closer to the other non-relativistic models; 
iii) the second derivatives of the curves are  associated
with the incompressibility and  can be very different. This is just a
reflex of 
the values tabulated for the compressibilities of the different models in
Tables \ref{tab:properties} and \ref{tab:propertiesrel}.

We next compare the isoscalar term of the interaction. In order to
separate the parabolic contribution we
represent this term divided by $\rho^2$.  The Walecka parametrization
shows an almost independent behavior on the density because this model
has no non-linear terms on the $\sigma$-meson. Non-relativistic and
relativistic models show similar behaviors although the last ones do
not grow so fast with density.  This tendency will be compensated by
the density dependence of the effective mass in the two types of models.
In fact, the effective mass  decreases faster in  the relativistic models  as 
compared with non-relativistic models, as seen in
Fig. \ref{sym}c). Smaller effective masses give rise to larger
kinetic energy contributions so that the kinetic energy contribution
to the total energy is more important in relativistic models. We
should stress, however that the effective masses in relativistic and
non-relativistic models have different meanings. For the first type the
effective mass includes 
the contribution of the nucleon scalar self-energy, while for the
second type the effective mass reflects the momentum dependence of the
single particle energy. We point out that the isoscalar channel of the
Skyrme parametrizations SLy230a and LNS are very similar. These two
parametrizations show differences in the isovector channel.

\begin{figure}[t]
\begin{center}
\begin{tabular}{cc}
\includegraphics[width=6.5cm,angle=0]{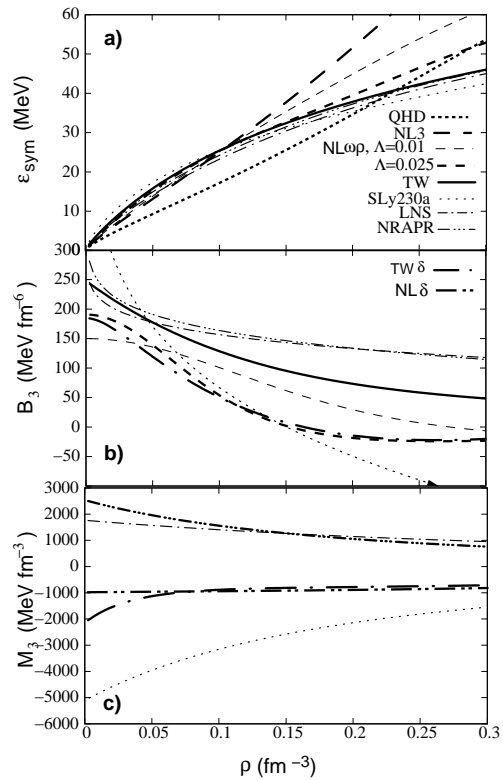} \\
\end{tabular}
\end{center}
\caption{Isospin channel: a) symmetry energy; b) exact $B_3$; c) isospin
  channel of effective mass $M_3$.}
\label{assym}
\end{figure}

Let us now turn to the isospin dependence. We focus the discussion on the symmetry 
energy, the  exact isovector term of the interaction $B_{3}$ [Eq. (\ref{b3exact})] and the
isospin channel of the effective mass $M_3$. We include the
relativistic models with non-linear terms on the $\rho-\omega$ mesons
which allow a more flexible description of the isovector channel. 
\noindent From Fig.  \ref{assym}a), one observes that the slopes of the symmetry energy
are somewhat different when considering a relativistic or a non-relativistic
model, TW being the only relativistic model that shows a behaviour similar to
the non-relativistic ones. The other relativistic models show an
approximately linear dependence of the  symmetry energy on the density.
 It has recently been discussed in \cite{bao-an} 
that there is a linear correlation between the neutron skin thickness and the 
slope of the symmetry energy for non-relativistic models and this fact gives a 
very strong constraint on the density dependence of the nuclear symmetry 
energy and consequently on the EoS as well. In \cite{skin07} it was found
that the same kind of correlation exists for density dependent relativistic 
models. 

The isovector channel of the interaction, Fig. \ref{assym}, also shows a large
discrepancy between the different models and also between models within
the same framework: in particular the SLy230a
shows a behavior very different for the other Skyrme parametrizations,
with a very fast decrease with density becoming even negative for
$\rho>0.15$ fm$^{-3}$. The relativistic models with the non-linear
$\omega-\rho$ terms or with density dependent couplings including the
$\delta$ meson also become negative at large densities but this effect is not
so pronnounced. 

In relativistic models there is a proton-neutron mass splitting only
if the scalar isovector $\delta$-meson is included. This occurs for
the DDH$\delta$ and NL$\delta$ parametrizations we consider. As we see
from Fig  \ref{assym}c), $M_3=(M^*_p-M^*_n)/\rho_3$ is negative for these models which
corresponds to a $M^*_n<M_p^*$ in neutron rich nuclear matter. A similar behavior is predicted by the
Skyrme interaction SLy230a  but an opposite behavior is obtained with
the LNS and NRAPR parametrizations of the Skyrme interaction. A discussion of the isospin
dependence of the effective mass has been done in \cite{baran05}.  The
proton-neutron mass splitting is  a present topic of discussion and
the forecoming experiments with  radioactive beams will allow the {clarification} of this 
point.  

From the figures just discussed we conclude that the isoscalar channel already shows 
some discrepancy between the
different models, mainly between the relativistic and the 
non-relativistic ones. However the larger discrepancies occur for the
isovector channel. In this case there is not even a general common
trend between the models of each class.

\section{Non-relativistic approximation to the energy functional}

In what follows we perform a non-relativistic expansion of the
energy density [Eq. (\ref{EQ:E})] in order to make a comparison with a classical approximation
like the Skyrme one. This approximation is based on the expansion
\begin{equation}
\epsilon_{i}^{*}=\sqrt{\mathbf{p}^{2}+{M_{i}^{*}}^{2}}\simeq {M_{i}^{*}+}\frac{%
\mathbf{p}^{2}}{2{M_{i}^{*}}}.  \label{Approxclass}
\end{equation}
In order to be valid the Fermi momentum should be small compared
to the nucleon effective mass meaning that such an expansion is
restricted to low densities. 
Then we have the scalar density and the relativistic kinetic energy 
written as functionals of the particle and kinetic densities: 
\begin{equation}
\rho _{s_{i}} \simeq \rho_{_{i}}-\frac{\tau _{i}}{2{M_{i}^{*^{2}}}},
\label{EQ:aproxrhos}
\end{equation}
\begin{equation}
K_{i} \simeq {M_{i}^{*}}\rho_{_{i}}+\frac{\tau _{i}}{2{M_{i}^{*}}}.
\label{EQ:aproxK}
\end{equation}
Using these two approximations we have reduced the relativistic densities 
to the standard densities appearing in the Skyrme approaches. Solving now the 
field equations it becomes easy to recast the relativistic energy functional 
in a Skyrme form for the considered case, depending only on particles 
and kinetic densities. Since this equivalence is obtained at the first order 
in $p^2/M^2$ we obtain analytical expressions for the fields at the same level 
of approximation. 
     
\subsection{\protect\smallskip QHD-II}

In order to test the non-relativistic approximation of the
relativistic models we first consider the QHD-II parametrization
\cite{qhd1}. In this case the non-linear terms are zero
($\kappa=0$, $\lambda=0$, and $\xi=0$) and the $\delta$ meson is not present
($c_{\delta}=0$). 

Then, the $\omega$ field is directly deduced from Eq. (\ref{EQ:omega})
\begin{equation}
{{V}}_{0} =g_{v} \rho/m_{v}^{2},
\label{V0QHD} 
\end{equation}
and so is the $\rho$ field from Eq. (\ref{EQ:rho})
\begin{equation}
b_{0} = \frac{g_{\rho }}{2} \rho_3 / m_{\rho }^{2}. 
\label{EQ:b0QHD} 
\end{equation}

In the same spirit,  the $\sigma$ field is directly deduced from 
Eq. (\ref{Eq:sigma}) leading to
\begin{equation}
\phi _{0}=g_{s} \rho_s/m_{s}^{2}.   \label{Eq:sigmaQHD}
\end{equation}
Using the expansion (\ref{EQ:aproxrhos}) of the scalar density 
$\rho_s$,  the $\sigma$ field
can be explicitly written as   a functional of $\rho$ and $\tau$
\begin{equation}
g_{s}{\phi}_{0}=2c_{s}(\rho -\frac{%
\tau }{2{M^{*^{2}}}}).
\label{phi0QHD}
\end{equation}
Since in this model the $\sigma$ field is the only contribution to
the effective mass, using (\ref{meff}),  
the effective mass is  given by:
\begin{equation}
M^{*}=M-2c_{s}(\rho - \frac {\tau}{{2M^*}^2}). \label{maqhd}
\end{equation}
This equation can be solved iteratively replacing $1/{M^*}^2$ on the rhs of 
equation (\ref{maqhd}) by $M^{-2} + 4 c_s \rho M^{-3}$.
The different terms of the energy (\ref{EQ:E}) are now all functional of
the particle densities and kinetic densities.

\begin{eqnarray}
 K_i &=& \left[ M-2c_{s}\left(\rho-\frac
    {\tau}{{2M^*}^2}\,\right)\right]\rho_i + \frac{\tau_i}{2{M^{*}}},\\
E_{\sigma } &=&c_{s}\rho ^{2}- c_{s}\frac {\tau}{{M^*}^2}\rho,\\
E_{\omega } &=&c_{v}\rho ^{2},\\
E_{\rho } &=&\frac{m_{\rho }^{2}}{2}b_{0}^{2}=c_{\rho }\rho
_{3}^{2},
\end{eqnarray}
and the energy functional can be analytically compared with
the Skyrme one, the differences being both in the functional dependence and in 
the coefficients. Then grouping the $K_i$ and $E_{\sigma }$ terms 
helps to recognize the different terms of the Skyrme functional: 
\[
\sum_i K_i+E_{\sigma }=\sum_{i=p,n}\left({M}{\rho
}_{i}{+T}_{i}\right) +E_{\sigma }^{\prime },
\]
with
\begin{eqnarray*}
T_{i} &=&{\frac{\tau_{i}}{2{M^{*}}}}, \\
E_{\sigma }^{\prime } &=&{-c_{s}\rho }^{2}, \\
\end{eqnarray*}
where the effective mass is approximated by Eq.
(\ref{maqhd}).

We can now look at the binding energy density given by
Eq.(\ref{bindef}) which becomes
\smallskip
\begin{equation}
B_{\rm non-rel.}=\sum_{i=p,n}{T}_{i}+E_{\sigma }^{\prime
}+E_{\omega }+E_{\rho }.
 \label{wm0}
\end{equation}
In order to test this Skyrme-like approximation, we rewrite the
functional $B_{\rm non-rel.}$ in the form of Eq. (\ref{skyr}) and
obtain
\begin{eqnarray}
B_{1}(\rho) &=&(c_{v}-c_{s})\rho^{2}
  \label{B1}, \\
B_{3}(\rho) &=& c_\rho
  \label{B3}.
\end{eqnarray}
In Figs. \ref{Bb1m}a) we compare the approximate binding energy,
$B_{\rm non-rel.}$, Eq.  (\ref{wm0}), with its exact value, Eq. 
(\ref{bindef}), the quantities $B_{1}(\rho)$, eq (\ref{B1}) with
$B_{1_{\rm exact}}(\rho)$ defined in Eq.  (\ref{b1exact}) and
$M^*$, Eq.  (\ref{maqhd}) with $M^*_{\rm \rm exact}=M-g_s \phi_0$. 
In all cases we can see that the non-relativistic approximation 
is a good approximation for low densities when the Fermi momentum 
is not too high. Thus, we can safely compare the functional obtained in
this limit with the Skyrme one.  

We can see that the above non-relativistic limit of this first simple 
relativistic model 
(QHD-II) leads to a Skyrme functional as far as the density dependence 
is concerned (see Eqs. (\ref{B1S}) and (\ref{B3S}) ), with a simple two body
 force ($C_3$ and $D_3$ are both zero).
As already referred, we point out that the three-body term in the Skyrme parametrization
  is essential to get saturation. For the simple Walecka model the quenching of the scalar
  field will play a similar role.  
The effective mass is similar but not identical since the Skyrme 
parametrization Eq.  (\ref{masseffS}) applies to $M^{*^{-1}}$ while the 
relativistic models introduce
directly  $M^*$ (see Eq.  (\ref{maqhd})). Moreover the relativistic models in 
their non-relativistic limit present a richer parametrization of the 
effective mass with non linear contributions in the kinetic energy density. 
However, the leading orders are similar. 

It should be noticed that, for this simple model, without a $\delta$ field the
isospin dependence of the effective mass is not introduced. In the next
section we  study the isospin dependence of the mass.    

In order to introduce non-trivial density dependence, 
we also consider models with interacting fields such as 
the non-linearities in the $\sigma$ models  or
the non-linear $\sigma \rho $ and $\omega \rho $
couplings (see section \ref{NLSROR}). An alternative way is to 
directly introduce density dependent coupling parameters as discussed in 
section \ref{DDCP}.

\subsection{\protect\smallskip $\delta$ and non-linearities in the $\sigma$ and $\omega$ fields}

In this section, we cure the two problems encountered in the simple model discussed above introducing the $\delta$ field in order to produce an isospin dependent mass splitting between protons and neutrons and a non-linear $\sigma$ and $\omega$ field to modify the density dependence of the energy functional.  

The first step is to solve the field equations. Let us first start with the 
$\delta$ field which can be easily solved 
introducing the approximate scalar field  (\ref{EQ:aproxrhos}) in the field equation 
 (\ref{EQ:delta}). We obtain for the $\delta_3$ field
\begin{equation}
\delta _{3}=2\frac{c_{\delta }}{\,g_{\delta }}\left( \rho
_{s_{p}}-\rho
_{s_{n}}\right) =2\frac{c_{\delta }}{\,g_{\delta }}\rho _{3}-\Delta %
_{K3},
\end{equation}
with
\[
\Delta _{K3}=\frac{c_{\delta }}{\,g_{\delta }}(\frac{\tau _{p}}{{%
M_{p}^{*^{2}}}}-\frac{\tau _{n}}{{M_{n}^{*^{2}}}}).
\]
        
Turning now to the $\sigma$ field we have to solve the self-consistent problem
relating ${\phi}_{0}$ to the proton and neutron scalar fields (\ref{Eq:sigma})
 which can be computed as a function of the proton and neutron effective 
masses, Eq.  (\ref{EQ:aproxrhos}), which is a function of the  $\sigma$ field,  
Eq.  (\ref{meff}). In the case of small non-linear terms,  the $\sigma$ field 
can be perturbatively solved.

Indeed, in a simple and crude approximation, the $\kappa$ and
$\lambda$ terms are assumed to be small, and then from Eq. 
(\ref{Eq:sigma}) the leading term becomes:
\begin{equation}
g_{s}\bar{\phi}_{0}=2c_{s}(\rho _{s_{p}}+\rho _{s_{n}})=2c_{s}(\rho -\frac{%
\tau _{p}}{2{M_{p}^{*^{2}}}}-\frac{\tau _{n}}{2{M_{n}^{*^{2}}}}).
\label{phi0}
\end{equation}
This approximation can thus be introduced in the non-linear terms
to get 
$\phi _{0}=\bar{\phi}_{0}+ d\phi$ with, at the lowest order in the non linear 
terms,
\begin{equation}
d\phi= 
-\frac{1}{2}\frac{\kappa }{m_s^2}\bar{\phi}_{0}^{2}-\frac{1}{6}%
\frac{\lambda}{m_s^2} \bar{\phi}_{0}^{3}.
\label{dphi}
\end{equation}
This leads to
\[
\phi _{0}=2\frac{c_{s}}{g_{s}}\rho -\Delta _{\sigma }-\Delta %
_{K0},
\]
with the non-linear and kinetic contributions 
\begin{eqnarray*}
\Delta _{\sigma } &=&2\kappa \frac{c_{s}^{2}}{{m_s^2}g_{s}^{2}}\rho ^{2}+%
\frac{4\lambda }{3}\frac{c_{s}^{3}}{{m_s^2}g_{s}^{3}}\rho ^{3}, \\
\Delta _{K0} &=&\frac{c_{s}}{g_{s}}(\frac{\tau _{p}}{{M_{p}^{*^{2}}}}%
+\frac{\tau _{n}}{{M_{n}^{*^{2}}}}).
\end{eqnarray*}

The same strategy can be used to solve the $\omega$ field either directly from
its relation to the baryonic density (cf. Eq.  (\ref{EQ:omega})) or assuming
that at low density the non-linearities are small and thus solving Eq.  
(\ref{EQ:omega}) iteratively from the linear solution 
${\bar{V}}_{0} =g_{v} \rho/m_{v}^{2}$.     
Then at the first iteration we get 
\begin{equation}
V_{0} =\frac {g_{v} \rho}{ m_{v}^{2}}
- \frac {1}{6} \xi g_v^7 \frac{\rho^3}{ m_{v}^{8}}.
\label{V0}
\end{equation}

The expressions for $\phi_0$ and $\delta_3$ can now be used to
obtain the non-relativistic approximation for the effective mass in the form
of  Eq.
(\ref{nlmeff}), with
\begin{eqnarray}
M_{1} &=& -2 c_{s}\rho + g_{s}\Delta _{\sigma } +  g_{s}\Delta _{K_0 }  
\label{M1a}, \\
M_{3} &=& -2 c_{\delta }  +  g_{\delta}\Delta _{K_3 }/\rho_3
\label{M3a}.
\end{eqnarray}
This set of coupled equations can be solved directly or iteratively. At the 
first iteration the $(M^*_i)^{-2}$  reads
\begin{equation}
(M^*_i)^{-2}=M^{-2} + 4 c_s \rho M^{-3} + 4 c_\delta \tau_{3i}
\rho_3 M^{-3}. \label{m2invi}
\end{equation}

Having solved the equations for the fields and for the effective masses in terms of the particle and kinetic density we can now study the energy functional.

In the present case, the different terms of the energy read
\begin{eqnarray}
 K_{i} &=&\rho _{_{i}} \left( M-2c_{s}\rho -\tau _{3i}~2c_{\delta }\rho _{3}+g_{s}
\Delta _{\sigma }
 \right. \\
& & \left.+g_{s}\Delta _{K0}+ g_\delta\tau _{3i}\Delta_{K3}\right)+
\frac{\tau _{i}}{2{M_{i}^{*}}}, \\
E_{\sigma } &=&c_{s}\rho ^{2}-g_{s}\rho (\Delta _{\sigma }+\Delta %
_{K0}) \nonumber \\
& & + \frac{1}{6}\kappa \phi_0 ^{3}+\frac{1}{24}\lambda \phi_0 ^{4}, \\
E_{\omega } &=&c_{v}\rho ^{2} +c_{v4} \rho^4, \qquad c_{v4}=- \frac{2}{3} \xi c_v^4, \\
E_{\delta } &=&c_{\delta }\rho _{3}^{2}-g_{\delta }\rho _{3}\Delta %
_{K3}, \\
E_{\rho } &=&c_{\rho }\rho _{3}^{2}. \label{eapprox}
\end{eqnarray}
Then grouping the terms $K_{i}$, $E_{\sigma }$ and $E_{\delta }$
leads to the following simplification
\[
\sum_i K_{i}+E_{\sigma }+E_{\delta }=\sum_{i=p,n}\left({M}{\rho
}_{i}{+T}_{i} \right) {+E_{\sigma }^{\prime }+E_{\delta }^{\prime
}},
\]
with
\begin{eqnarray*}
T_{i} &=&{\frac{\tau _{i}}{2{M_{i}^{*}}}}, \\
E_{\sigma }^{\prime } &=&{-c_{s}\rho }^{2}+c_{s_{3}}\rho
^{3}+c_{s_{4}}\rho
^{4}, \\
E_{\delta }^{\prime } &=&{-}c_{\delta }\rho _{3}^{2},
\end{eqnarray*}
where the effective mass is given by equation (\ref{nlmeff}) with (\ref{M1a})
and (\ref{M3a}).  

The non-linear effects are included in the additional parameters
$c_{s_{3}}=4\kappa c_{s}^{3}/3g_{s}^{3}=\kappa g_{s}^{3}/6
m_s^{6}$ and $c_{s_{4}}=2\lambda c_{s}^{4}/3g_{s}^{4}=\lambda
g_{s}^{4}/24m_s^{8}$.

The binding energy density $B={E}-M\rho,$ reads now
\begin{equation}
B_{\rm non-rel.}=\sum_{i=p,n}{T}_{i}+E_{\sigma }^{\prime
}+E_{\omega }+E_{\delta }^{\prime }+E_{\rho }, \label{nlwm0}
\end{equation}
and we get for $B_1$ and $B_3$
\begin{eqnarray}
B_{1}(\rho ) &=&(c_{v}{-c_{s})\rho }^{2}+c_{s_{3}}\rho
^{3}+(c_{s_{4}}
+ c_{v_4}) \rho^{4}  \label{B1a}, \\
B_{3}(\rho ) &=&c_{\rho }{-}c_{\delta }  \label{B3a}.
\end{eqnarray}

As in the simplest QHD model, taking advantage of the non-relativistic limit
we have  expressed the scalar density and the relativistic kinetic energy
density as functionals of $\rho_i$ and $\tau_i$. Then the energy also becomes
a   functional of $\rho_i$ and $\tau_i$ of the form of a Skyrme
functional. The interaction part presents a very similar structure. 
In Figs. \ref{Bb1m}a), b) and c) we show how the approximation works for NL3,
TM1 and NL$\delta$. While for TM1 the present approximation is good, for NL3
and NL$\delta$ we would have to include higher orders in the density expansion
to improve the results.
The non-linear coupling in the $\sigma$ and $\omega$ fields have introduced
higher order terms in the potential energy of symmetric matter. However
because of the perturbative approach we have taken we are restricted to a
polynomial density dependence. To go beyond this limitation, we have also
fitted Skyrme parameters on the exact potential energy functional.  The
results of this fit is given in Table \ref{para1} and will be  discussed later.
As far as the isospin dependence is concerned, the absence of non linearities (or couplings) in isovector fields leads to a rather poor isospin dependence since $B_3$
is constant. The introduction of a coupling of the isoscalar fields with the isovector $ \rho $ field corrects this fact. 
Finally, the main difference is again in the functionals describing the effective mass but now not only the expansions are made for different quantities, the mass in relativistic approach and the inverse mass in the Skyrme model, but also the relativistic approaches lead to a much richer functional of both $\rho_i$ and $\tau_i$.

\subsection{\protect\smallskip Non-linear $\sigma \rho $ and $\omega \rho $
couplings}
\label{NLSROR}

Still a different model includes non-linear $\sigma-\rho$ and
$\omega-\rho$ couplings \cite{hor,bunta} which allow to change the
density dependence of the symmetry energy. In the corresponding
Lagrangian a new coupling term $\mathcal{L}_{{\sigma\omega\rho }}$
is added:
\begin{eqnarray}
\mathcal{L}_{{\sigma\omega\rho }} &=&g_{\rho }^{2}\vec{b}_{\mu
}\cdot \vec{b}^{\mu }[\Lambda _{s}g_{s}^{2}\phi ^{2}+\Lambda
_{v}g_{v}^{2}V_\mu V^{\mu }].
\end{eqnarray}
We have followed the prescription of \cite{hor} so that the coupling $%
\Lambda _{i}$ is chosen in such a way that for $k_{F}=1.15$
fm$^{-1}$ (not the saturation point) the symmetry energy is 25.68
MeV like in the NL3 parametrization. 
We start by setting $\Lambda
_{s}=0$ as in \cite{bunta2}. In this case, the equations of motion
are not the standard ones, once two of them become coupled and,
for this reason, they are reproduced as follows:
\begin{eqnarray*}
g_{v}V_{0} &=&\frac{g_{v}^{2}}{m_{v}^{2}}\left[ \rho
-2g_{v}\,V_{0}\,g_{\rho
}^{2}b_{0}^{2}\Lambda _{v}\right] , \\
\frac{g_{\rho }}{2}b_{0} &=&\frac{g_{\rho }^{2}}{4m_{\rho
}^{2}}\left[ \rho _{3}-4\,g_{\rho }b_{0}g_{v}^{2}V_{0}^{2}\Lambda
_{v}\right] .
\end{eqnarray*}
If the $\Lambda _{s}$ is not assumed to be zero but no
non-linearity is taken into account either in the $\sigma$ or in
the $\omega$ field (i.e. $\kappa =0$, $\lambda =0$ and $\xi=0$),
the three fields are coupled in the following way:

\begin{eqnarray*}
g_{s}\phi _{0} &=&\frac{g_{s}^{2}}{m_{s}^{2}}\left[ \rho
_{s}-2g_{s}\,\phi
_{0}\,g_{\rho }^{2}b_{0}^{2}\Lambda _{s}\right] , \\
g_{v}V_{0} &=&\frac{g_{v}^{2}}{m_{v}^{2}}\left[ \rho
-2g_{v}\,V_{0}\,g_{\rho
}^{2}b_{0}^{2}\Lambda _{v}\right] , \\
\frac{g_{\rho }}{2}b_{0} &=&\frac{g_{\rho }^{2}}{4m_{\rho
}^{2}}\left[ \rho _{3}-4\,g_{\rho }b_{0}(g_{s}^{2}\,\phi
_{0}^2\,\Lambda _{s}+g_{v}^{2}V_{0}^{2}\Lambda _{v})\right] .
\end{eqnarray*}
This set of equations should be solved self consistently. However,
if the $\Lambda_s$ and $\Lambda_v$
are small we can solve the problem perturbatively introducing $%
\phi _{0}=\frac{g_{s}}{m_{s}^{2}}\rho_{s} $ ,
$V_{0}=\frac{g_{v}}{m_{v}^{2}}\rho $ and $b_{0}=\frac{g_{\rho
}}{2m_{\rho }^{2}}\rho _{3}$ in the right hand side of the above
equations
\begin{eqnarray*}
g_{s}\phi _{0} &=&2c_{s}\rho _{s}\left[1-64c_{s}c_{\rho }^{2}
\,\rho
_{3}^{2}\Lambda _{s}\right] , \\
g_{v}V_{0} &=&2c_{v}\rho \left[ 1-64c_{v}c_{\rho }^{2}\,\rho
_{3}^{2}\Lambda
_{v}\right] , \\
\frac{g_{\rho }}{2}b_{0} &=&2c_{\rho }\rho _{3}\left[
1-64\,c_{\rho }(c_{s}^{2}\rho_{s}^{2}\,\Lambda _{s}+c_{v}^{2}\rho
^{2}\Lambda _{v})\right] .
\end{eqnarray*}
then the binding energy reads

\begin{equation}
B=\sum_{i=p,n}{T}_{i}+E_{\sigma }^{\prime }+E_{\omega }+E_{\delta
}^{\prime}+ E_{\rho }+E_{\rho sv},
\end{equation}
where the last term is the interaction energy between the $\rho $
and the $\sigma $ and $\omega $ fields
\[
E_{\rho sv}=g_{\rho }^{2}b_{0}^{2}[\Lambda _{s}g_{s}^{2}\phi
_{0}^{2}+\Lambda _{v}g_{v}^{2}V_{0}^{2}].
\]

\begin{figure}[t]
\begin{center}
\begin{tabular}{cccccc}

\includegraphics[width=5.5cm]{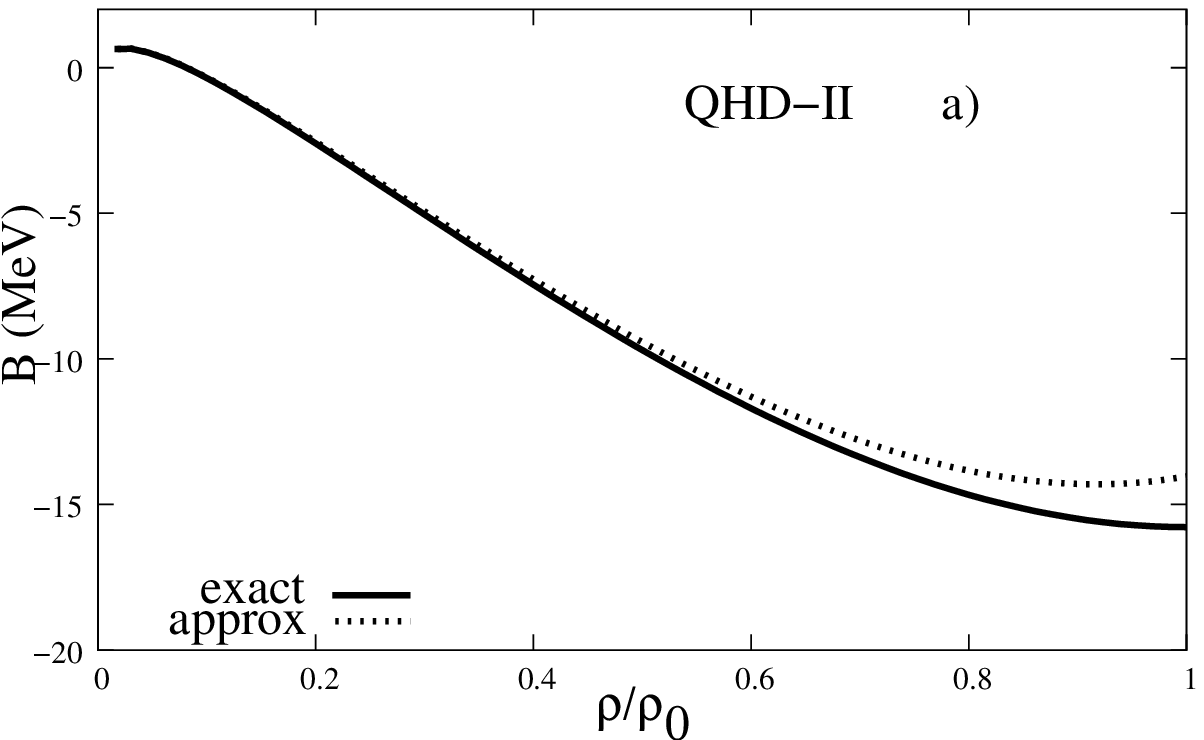} &
\includegraphics[width=5.5cm]{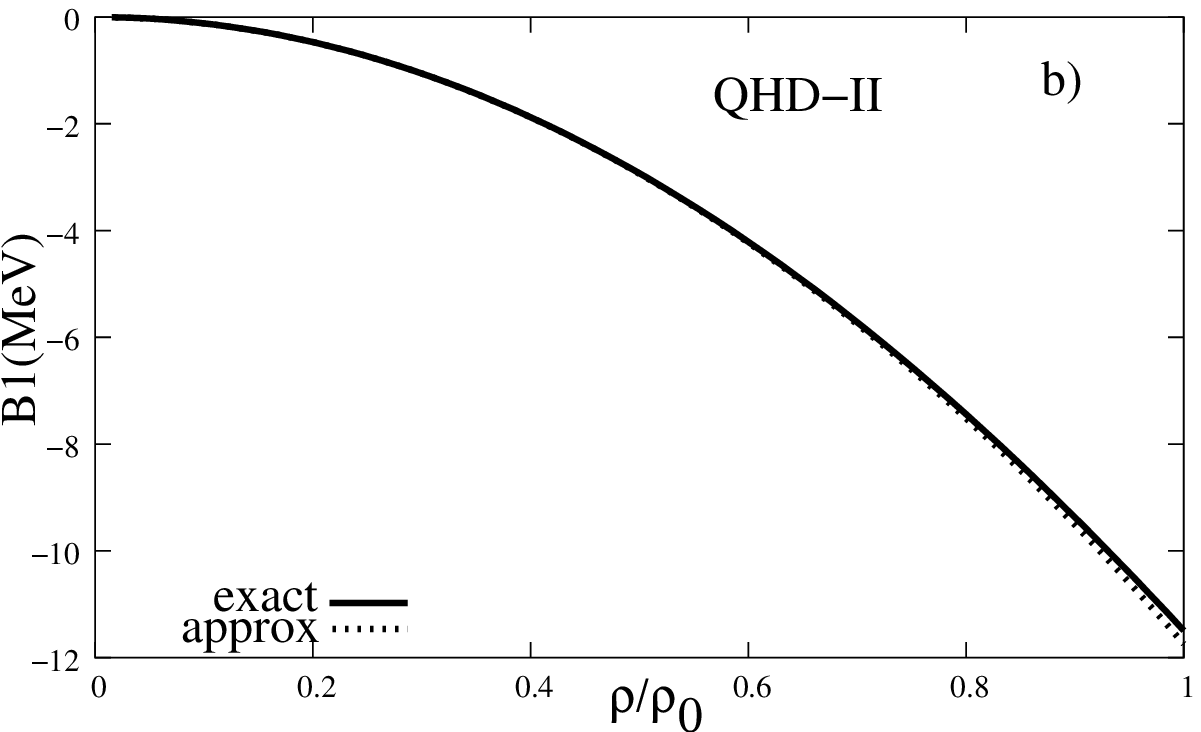} &
\includegraphics[width=5.5cm]{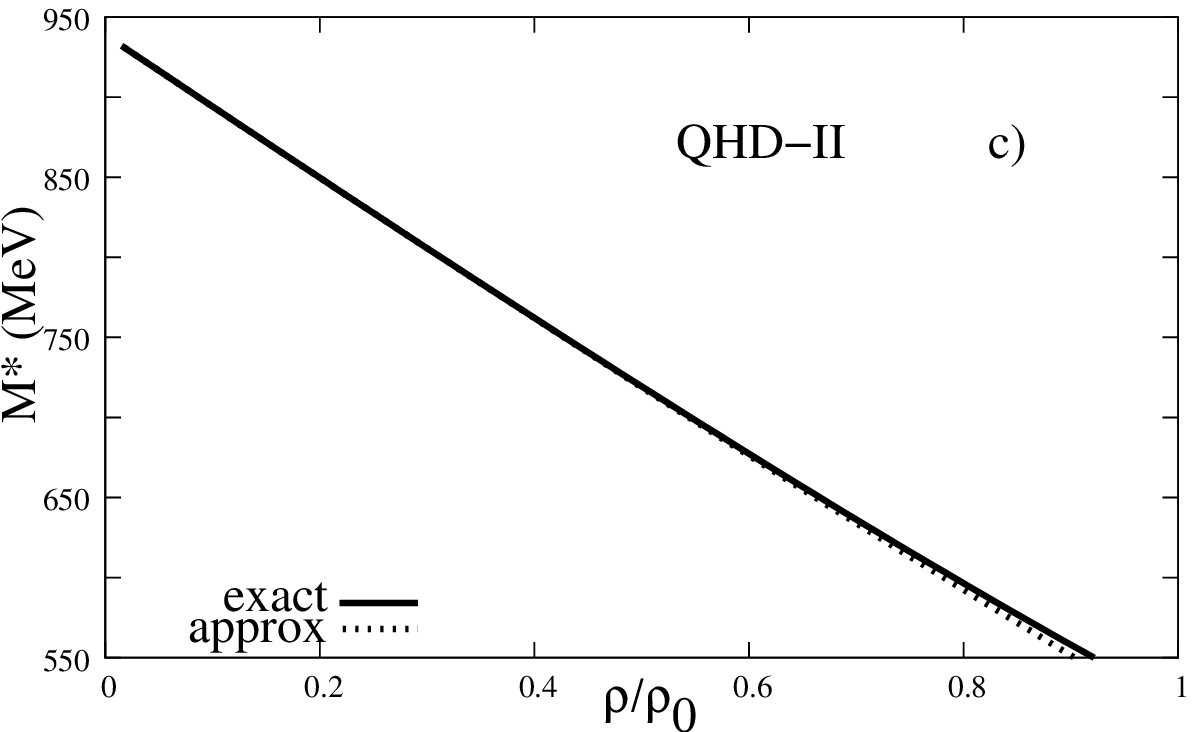}\\

\includegraphics[width=5.5cm]{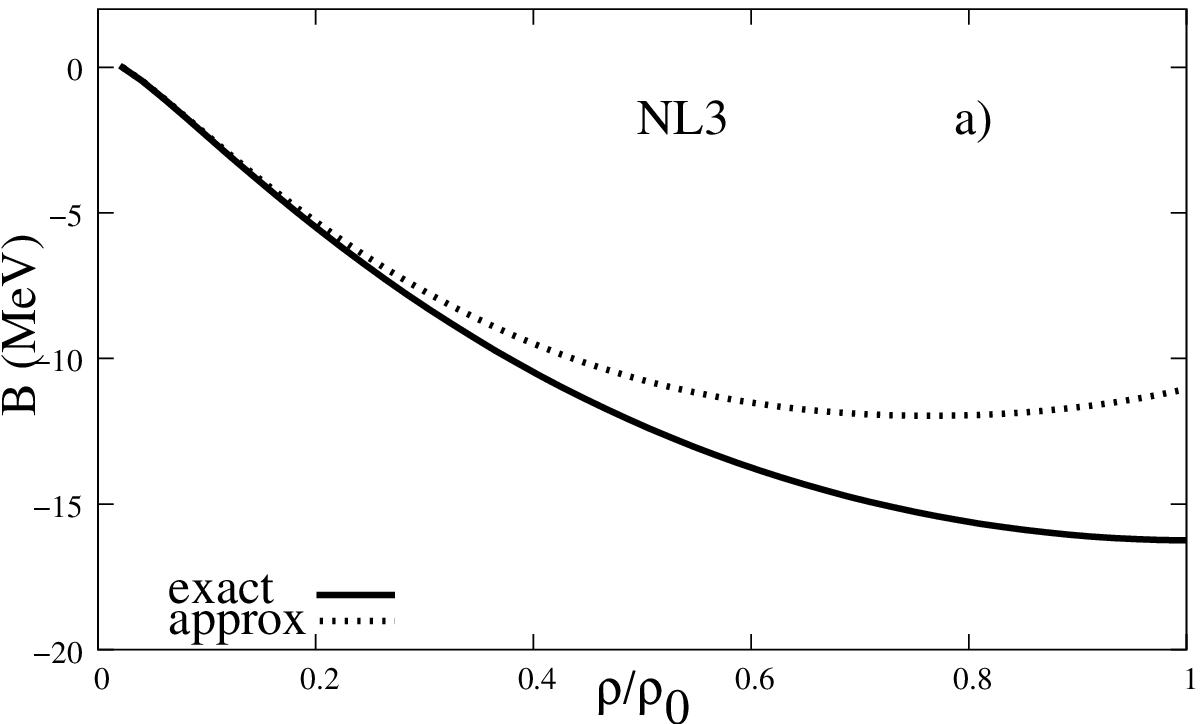} &
\includegraphics[width=5.5cm]{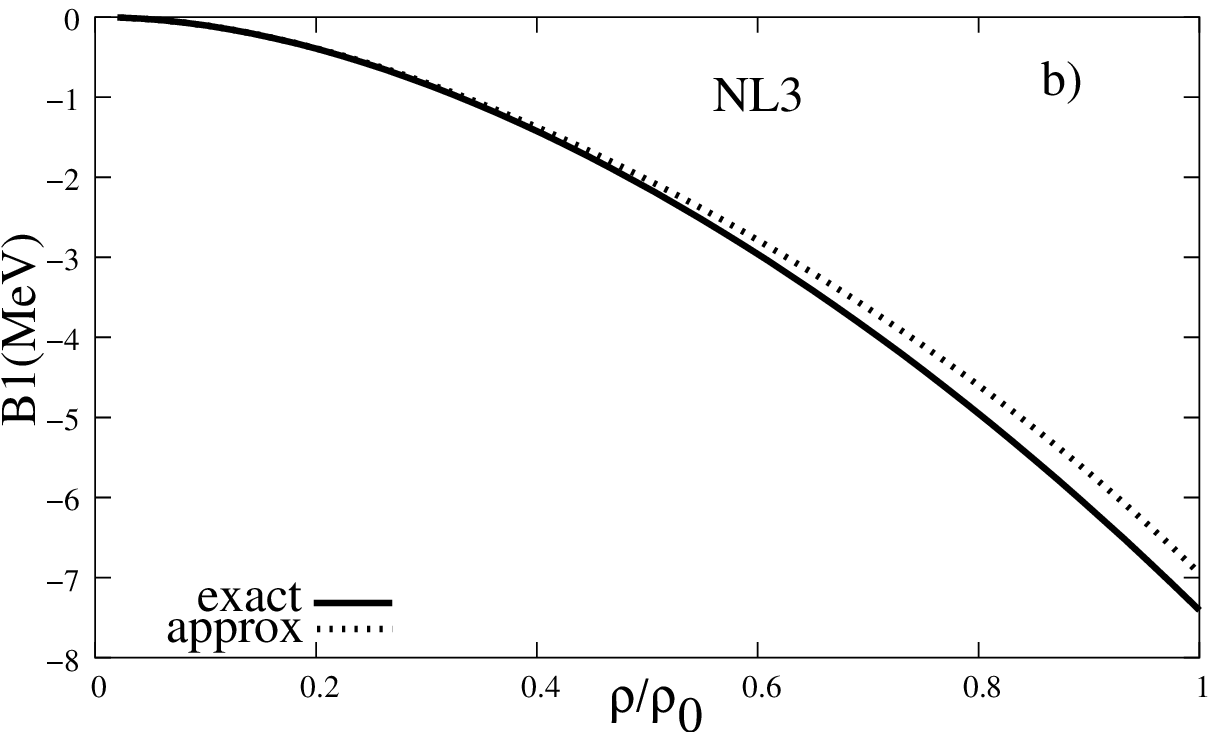} &
\includegraphics[width=5.5cm]{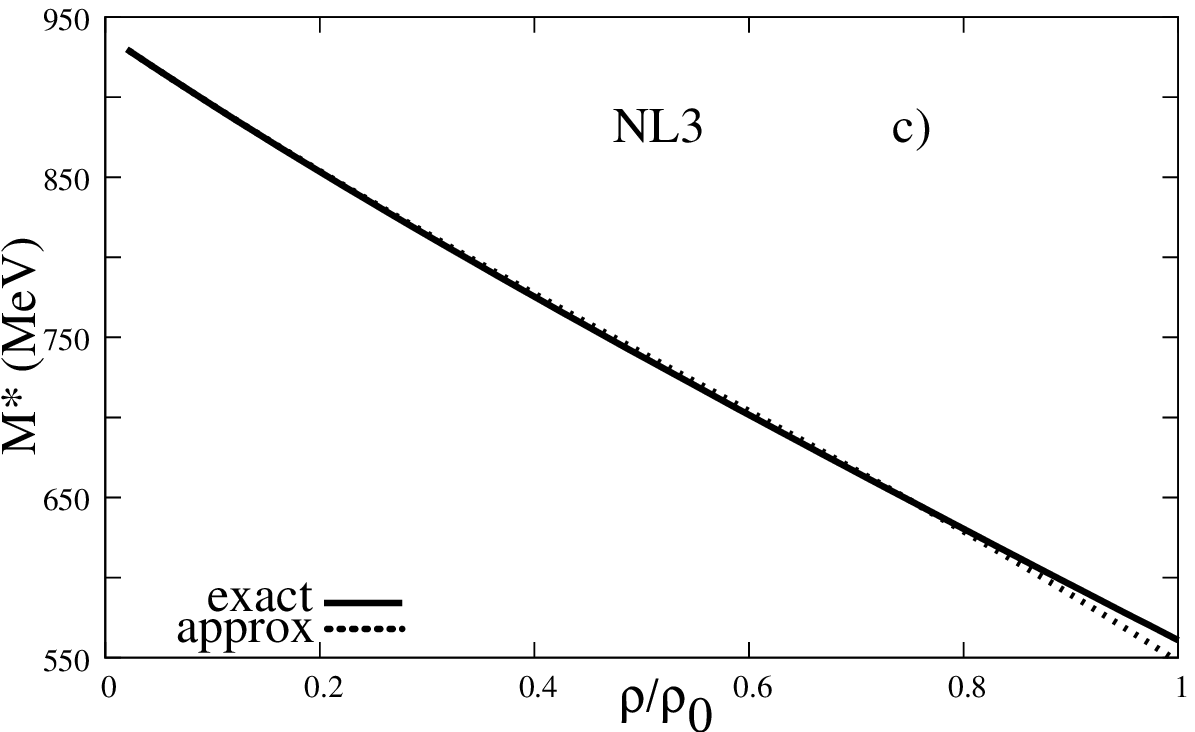}\\

\includegraphics[width=5.5cm]{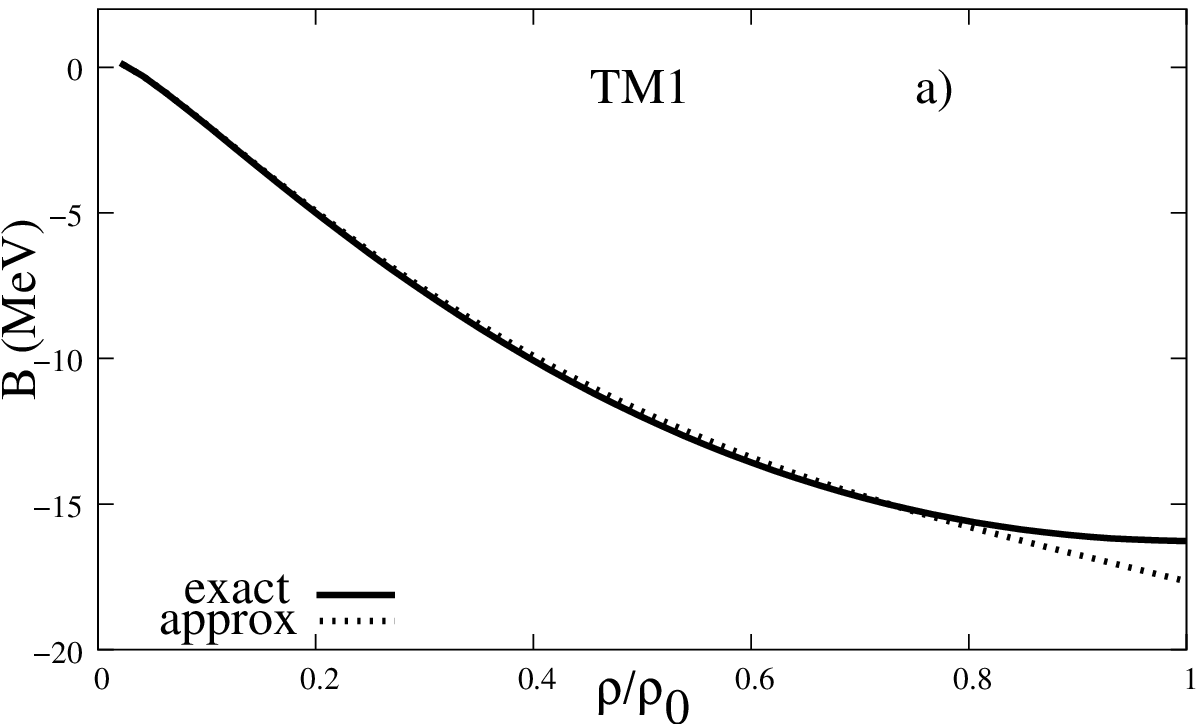} &
\includegraphics[width=5.5cm]{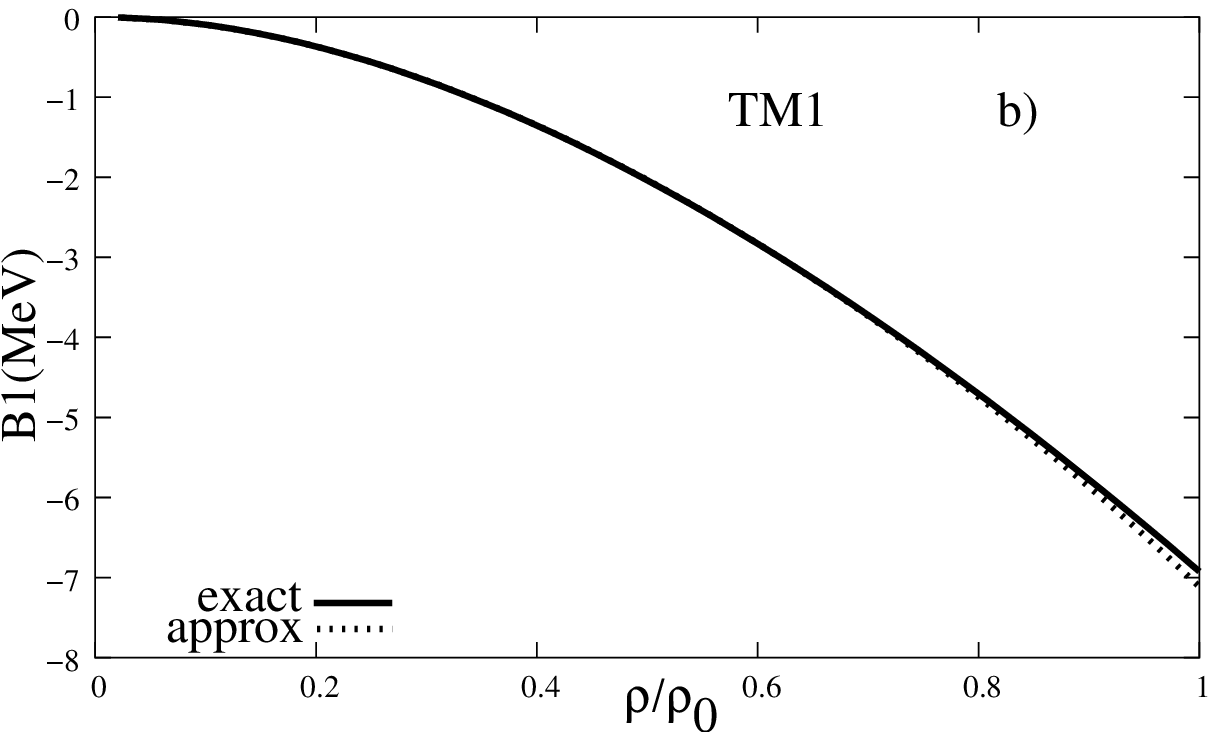} &
\includegraphics[width=5.5cm]{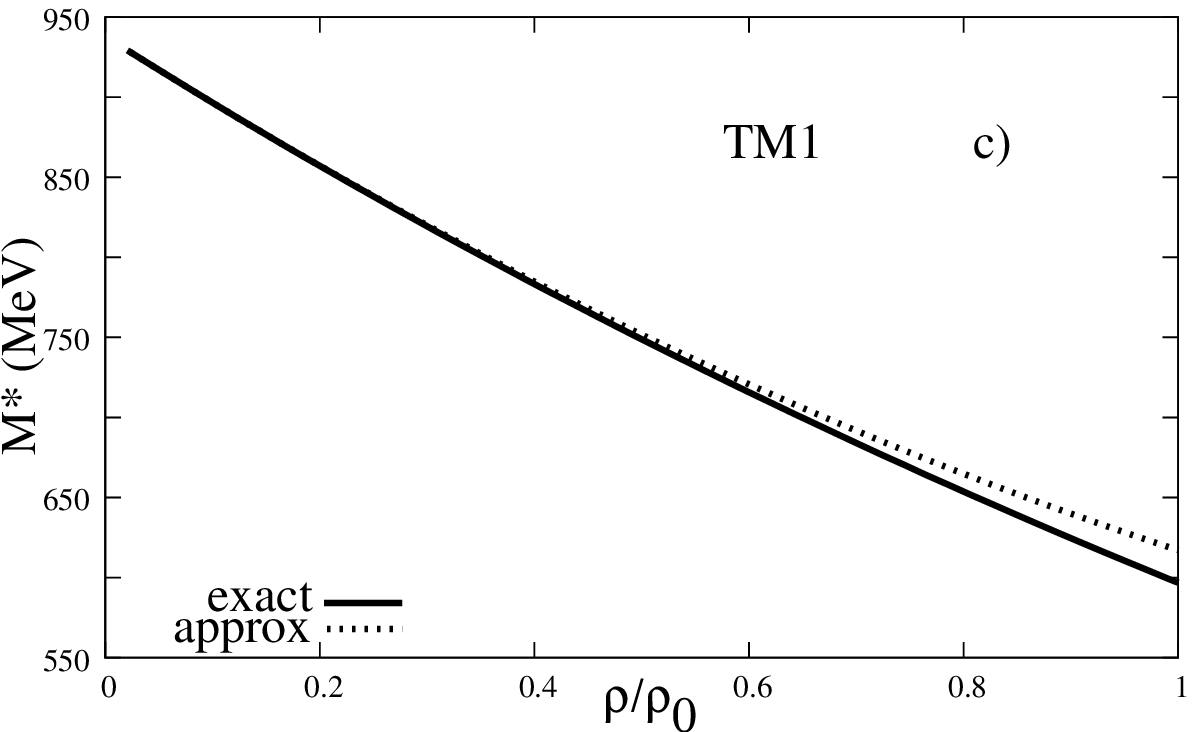}\\

\includegraphics[width=5.5cm]{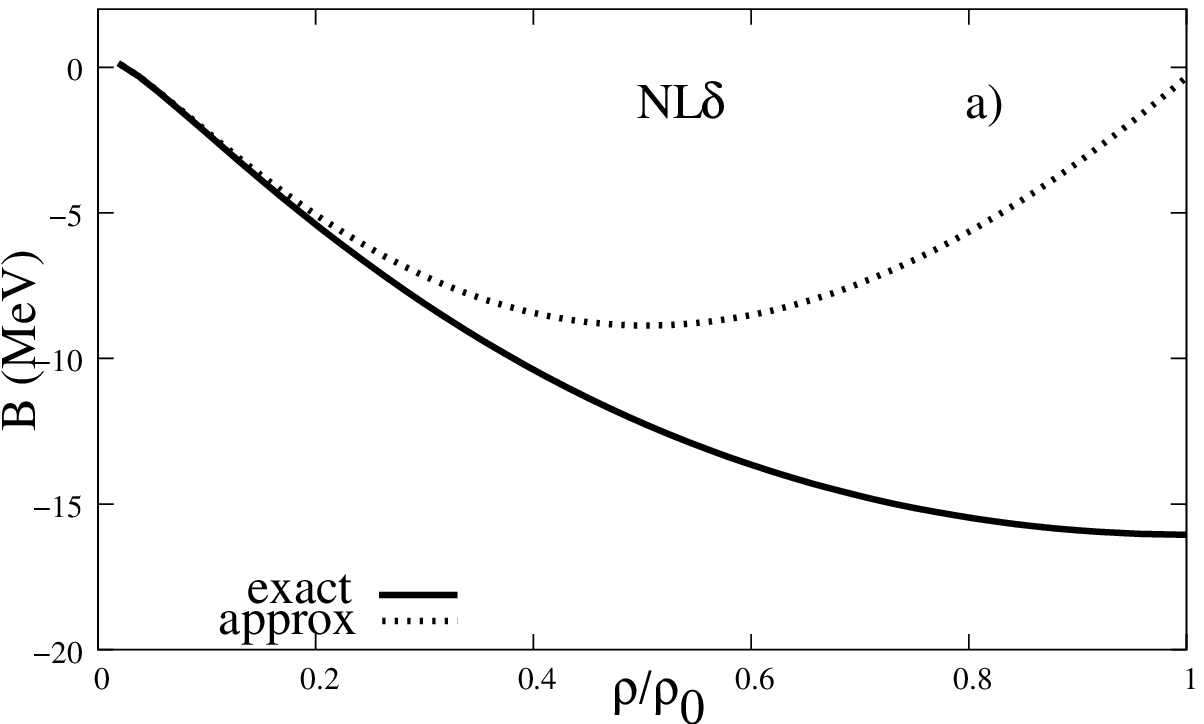} &
\includegraphics[width=5.5cm]{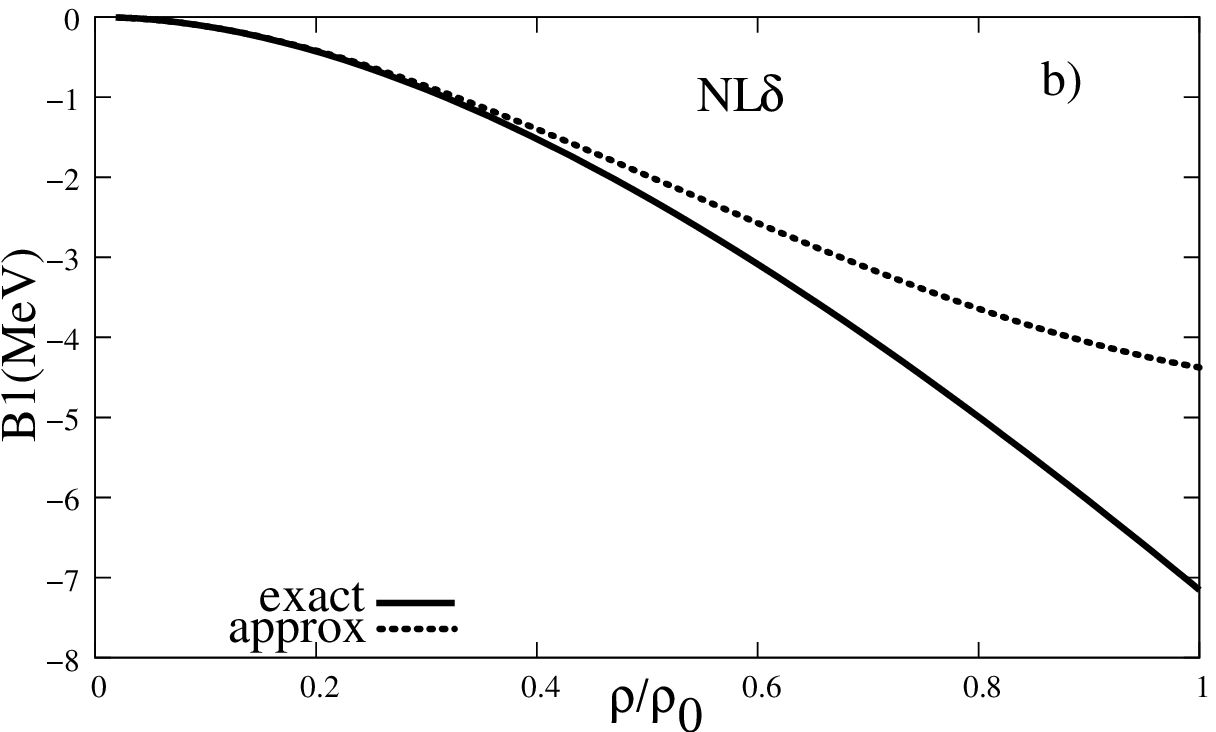} &
\includegraphics[width=5.5cm]{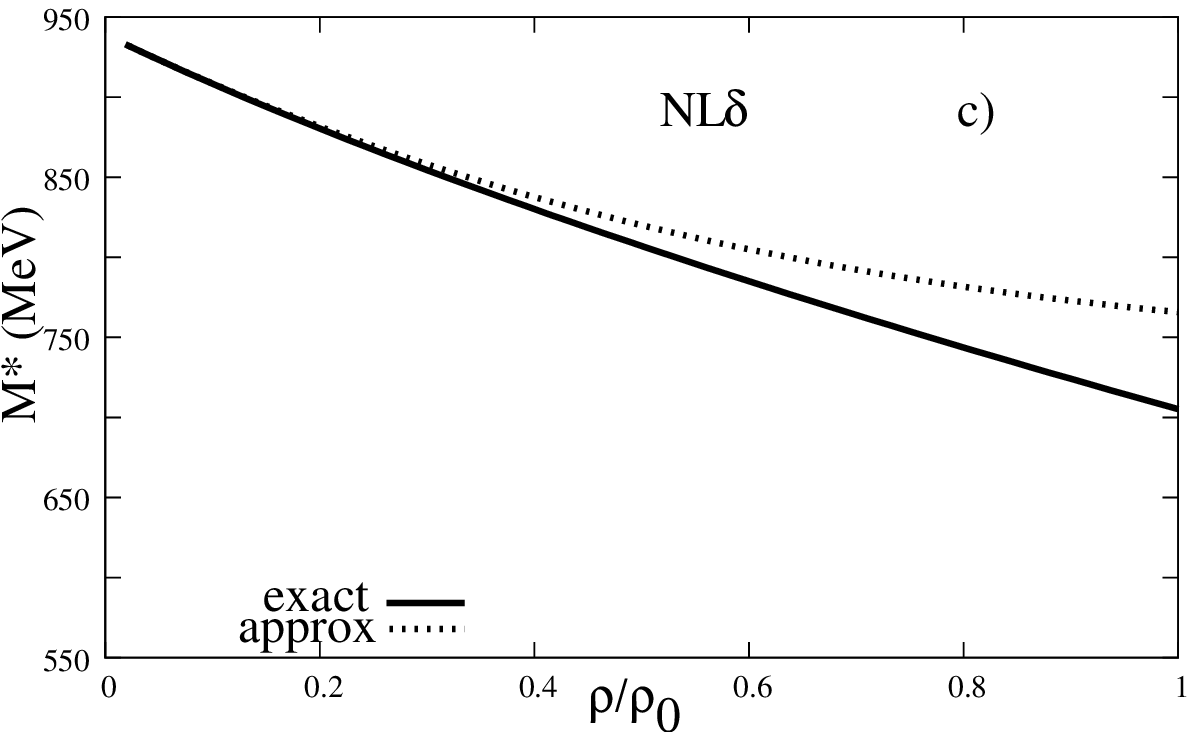}\\

\includegraphics[width=5.5cm]{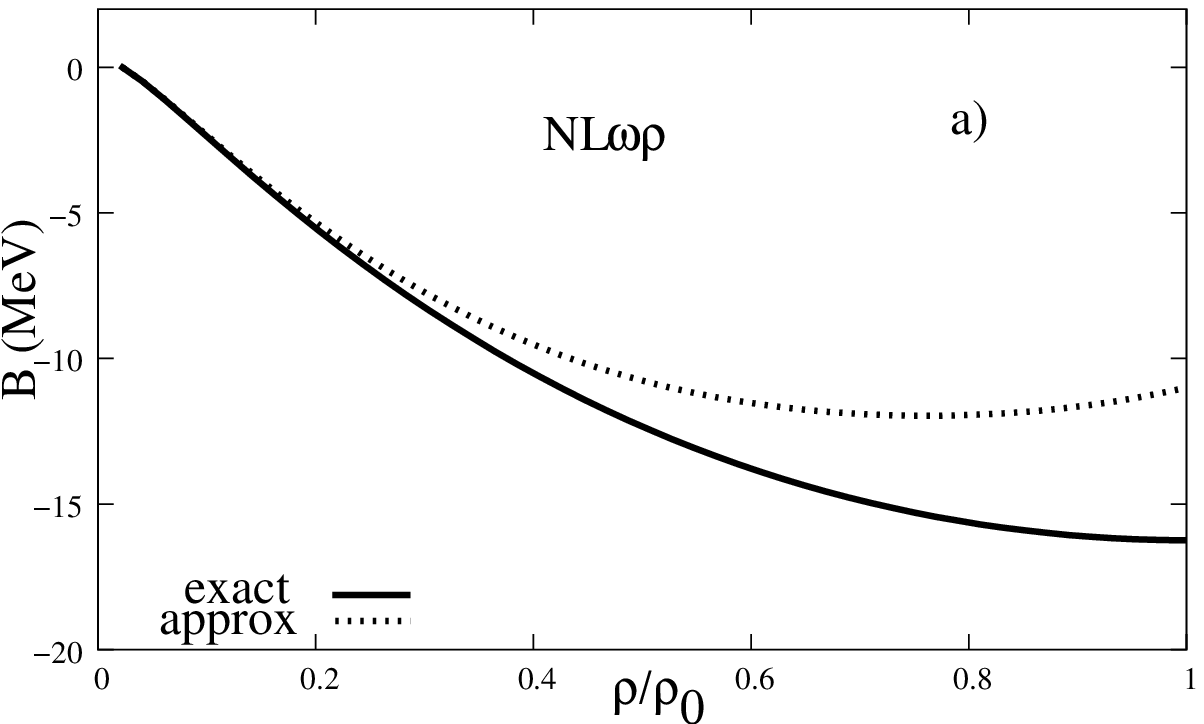} &
\includegraphics[width=5.5cm]{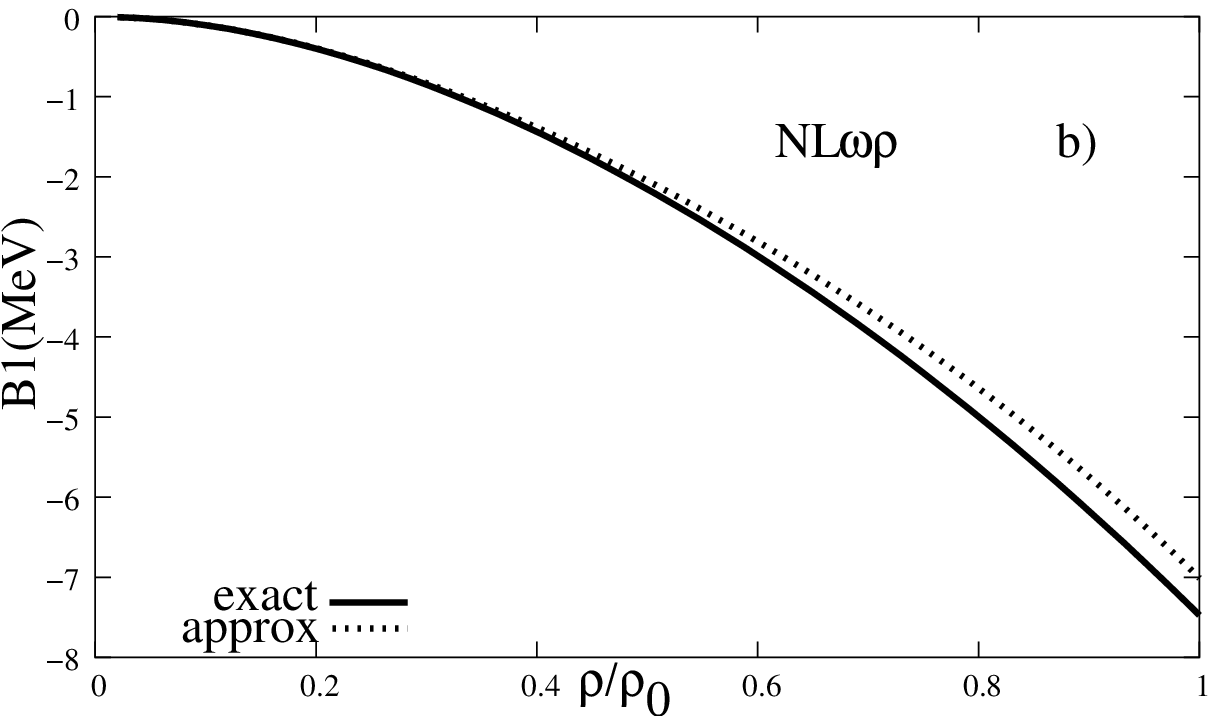} &
\includegraphics[width=5.5cm]{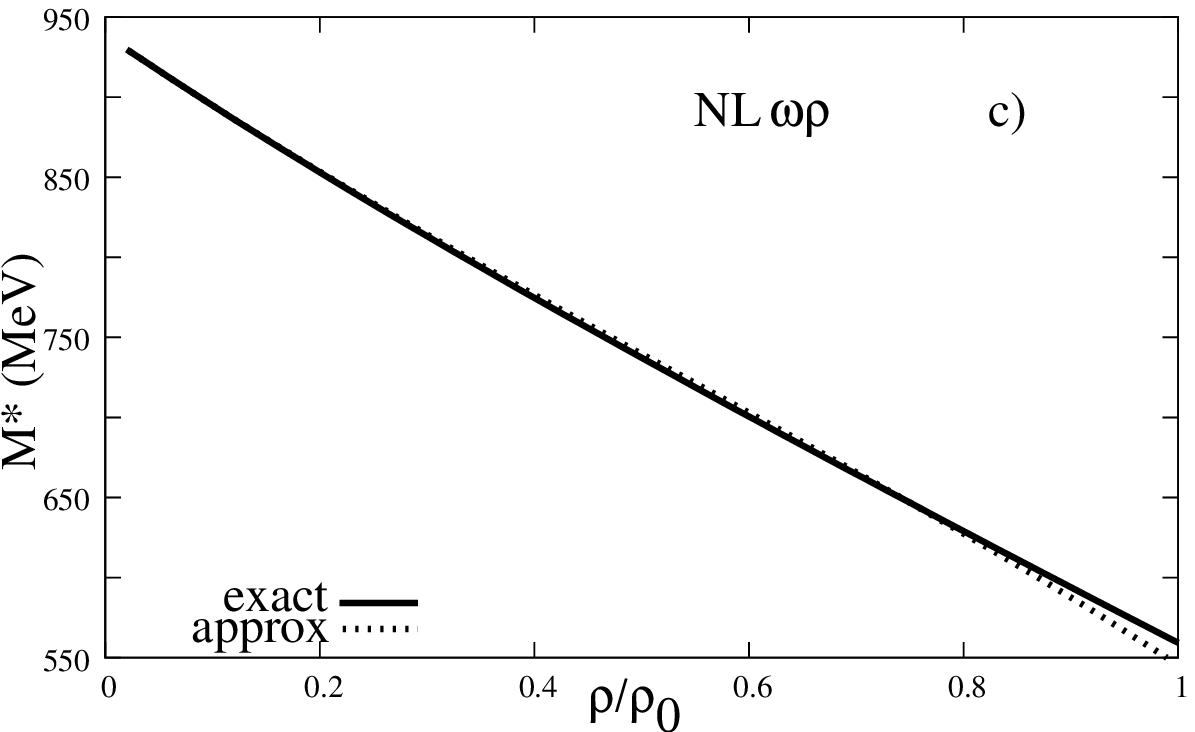}\\

\end{tabular}
\end{center}

\caption{Comparison between several exact and approximate physical quantities:
  a) binding energy density, b) $B_1(\rho)$ coefficient and
c) effective mass $M^*$ for relativistic models with constant
couplings.}\label{Bb1m}
\end{figure}

In this case,
\begin{eqnarray}
T_{i} &=&{\frac{\tau _{i}}{2{M_{i}^{*}}}}, \\
E_{\sigma }^{\prime } &=&{-c_{s}\rho }^{2},\\
E_{\omega } &=&c_{v}\rho ^{2}-2c_{v\rho }\,\rho ^{2}\rho _{3}^{2}, \\
E_{\delta }^{\prime } &=&{-}c_{\delta }\rho _{3}^{2}, \\
E_{\rho } &=&c_{\rho }\rho _{3}^{2}-2c_{v\rho }\,\rho ^{2}\rho
_{3}^{2}-2c_{s\rho }\,\rho ^{2}\rho _{3}^{2}, \\
E_{\rho sv} &=&c_{s\rho }\,\rho ^{2}\rho _{3}^{2}+c_{v\rho }\,\rho
^{2}\rho _{3}^{2},
\end{eqnarray}
with $c_{v\rho }=64c_{v}^{2}c_{\rho }^{2}\Lambda _{v},$ $c_{s\rho
}=64c_{\rho }^{2}c_{s}^{2}\,\Lambda _{s}.$ The effective mass can
again be approximated by the leading terms of equation
(\ref{nlmeff}). It should be noticed that the last term of $E_{\rho sv}$ partly
cancels the $E_{\rho }$ correction leading to
$$
E_{\rho }+E_{\rho sv}=c_{\rho }\rho_{3}^{2}-c_{v\rho}\,\rho ^{2}
\rho_{3}^{2}-c_{s\rho }\rho ^{2}\rho_{3}^{2}.
$$
The $B$ coefficients are then written as
\begin{eqnarray}
B_{1}(\rho ) &=&(c_{v}-c_{s})\rho^{2}+c_{s_{3}}\rho
^{3}+c_{s_{4}}\rho
^{4} \\
B_{3}(\rho ) &=& c_{\rho}-c_{\delta }-(3c_{v\rho }\,+c_{s\rho
})\rho ^{2},
\end{eqnarray}
where we also include the contributions of the non-linear $\sigma$
terms. These expressions increase more rapidly with density than
the corresponding Skyrme functional contribution given by
equations (\ref{B1S}-\ref{B3S}): for the relativistic model a 
term with a $\sigma=2$ exponent would be 
necessary while the usual range of the Skyrme parameter $\sigma$ is below 1 in
order to not present a too strong incompressibility.
In Fig. \ref{Bb1m} we see that the present approximation works well for $B_1$
and the
effective mass $M^*$ but fails  to give a reasonable description of the
binding energy. This was expected because NL$\omega\rho$ is just NL3 with non
linear $\omega\rho$ terms. The limitation of the present approximation is also
clear for the $B_3$ term shown in Fig. \ref{B33rm}a).
 Since we are looking at a
low density expansion we have also directly fitted Skyrme parameters on the
relativistic symmetry energy, Table \ref{para1}.

\subsection{Density dependent coupling parameters}
\label{DDCP}

Next we consider two models with density dependent coupling
parameters, respectively the TW model \cite{TW} and the DDH$\delta
$ \cite{gaitanos,inst} which also includes the $\delta $ meson.
These two models do not include self-interaction terms for the
meson fields (i.e. $\kappa =0$, $\lambda =0$ and $\xi=0$ ). The
only difference comes from the replacement of $g$ coupling
constants for the density dependent coupling parameters
$\Gamma_{s}$, $\Gamma_{v}$, $\Gamma_{\rho }$ and $\Gamma_{\delta
}$ which are adjusted in order to reproduce some of the nuclear
matter bulk properties, using the following parametrization for
the TW model
\begin{equation}
\Gamma _{i}(\rho )=\Gamma _{i}(\rho _{sat})h_{i}(x),\quad x=\rho
/\rho _{sat}, \label{paratw1}
\end{equation}
with
\begin{equation}
h_{i}(x)=a_{i}\frac{1+b_{i}(x+d_{i})^{2}}{1+c_{i}(x+d_{i})^{2}},\quad
i=s,v
\end{equation}
and
\begin{equation}
h_{\rho }(x)=\exp [-a_{\rho }(x-1)],  \label{paratw2}
\end{equation}
and $\Gamma _{\delta }(\rho )=0$, with the values of the
parameters $m_{i}$, $\Gamma _{i}$, $a_{i}$, $b_{i}$, $c_{i}$ and
$d_{i}$, $i=s,v,\rho $ given in \cite{TW}. \noindent For the
DDH$\delta $ model we consider the TW parametrizations of $\Gamma
_{s}$ and $\Gamma _{v}$ and for the other two mesons we take
\[
h_{i}(x)=a_{i}\exp [-b_{i}(x-1)]-c_{i}(x-d_{i}), \qquad i=\rho
,\,\delta .
\]
Such density dependences in the coupling parameters do not affect
the expression for the energy functional but of course affect its
derivative such as the pressure or the chemical potentials. The
latter ones are given by
\begin{equation}
\mu_i=\nu_i + \Gamma_{v} V_0 +\tau_{i3} \frac{\Gamma_{\rho}}{2}
b_0 + \Sigma_0^R,  \label{mu}
\end{equation}
where the rearrangement term is
$$
\Sigma_0^R=\frac{\partial\, \Gamma_v}{\partial \rho}\, \rho\,
V_0+\frac{\partial\, \Gamma_\rho}{\partial \rho}\, \rho_3\,
\frac{b_0}{2}- \frac{\partial\, \Gamma_s}{\partial \rho}\,
\rho_s\, \phi_0 -\frac{\partial\, \Gamma_\delta}{\partial \rho}\,
\rho_{s3}\, \delta_3.
$$

\begin{figure}
\begin{center}
\begin{tabular}{cc}
\includegraphics[width=6.5cm,angle=0]{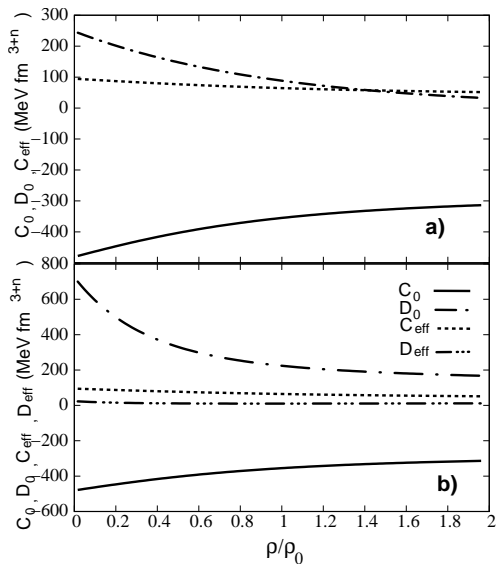}\\
\end{tabular}
\end{center}
\caption{Coefficients $C_i$ and $D_i$ for (a) the TW model and (b)
the DDH$\delta$ model.} \label{coef}
\end{figure}

As already discussed in the literature \cite{inst,br,spinodal},
the rearrangement term is crucial in obtaining different behaviors
in many quantities related to the chemical potentials or to their
derivatives with respect to the density as compared with the more
common NL3 or TM1 parametrizations.

The binding energy functional is given by equation (\ref{nlwm0})
with

\begin{eqnarray}
B_{1}(\rho ) &=&(C_{v}-C_s)\rho^{2}=C_0 \rho^{2}, \label{b1tw}\\
B_{3}(\rho ) &=&C_{\rho }-C_{\delta}= D_0, \label{b3tw}\\
M_{i}^{*} &=&M-2C_{s}\rho -\tau_{3i}~2C_{\delta }\rho _{3},\label{mtw}\\&=&M-2C_{\rm eff}\rho -\tau_{3i}~2D_{\rm eff }\rho _{3},
 \nonumber\end{eqnarray} where the $C_{i}$ coefficients are
computed replacing the coupling constants $g_{i}$ by $\Gamma
_{i}(\rho)$.

\begin{figure}[t]
\begin{center}
\begin{tabular}{cccccc}

\includegraphics[width=5.5cm]{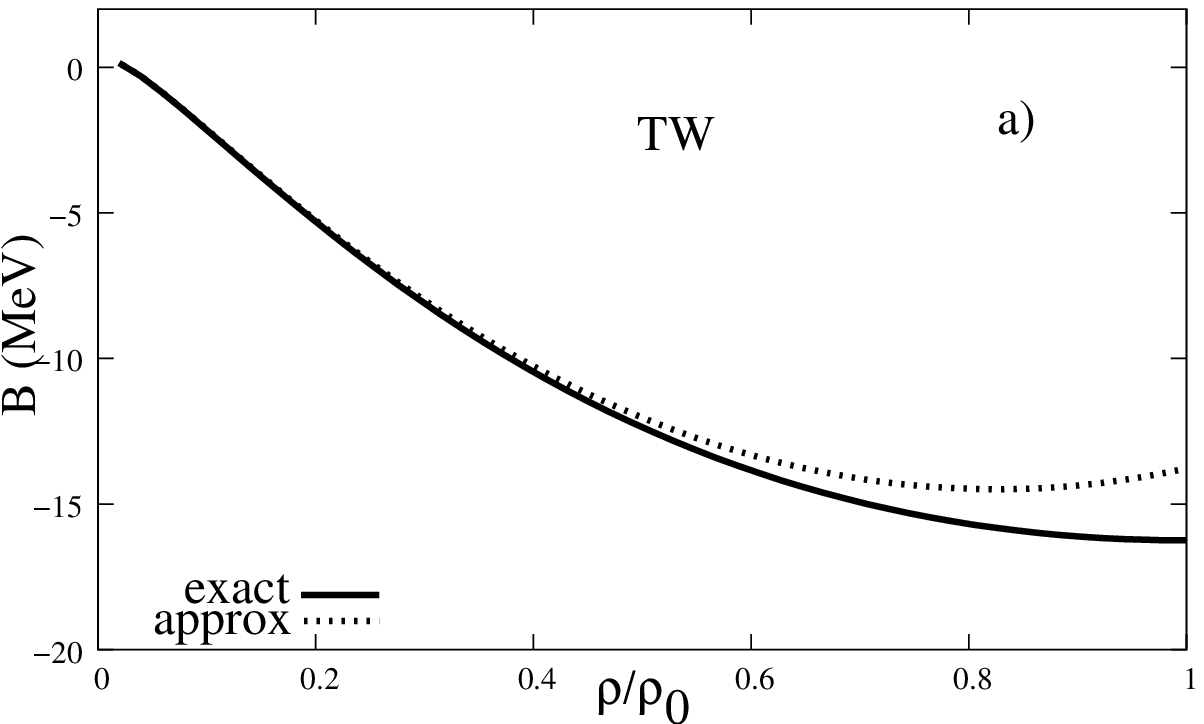} &
\includegraphics[width=5.5cm]{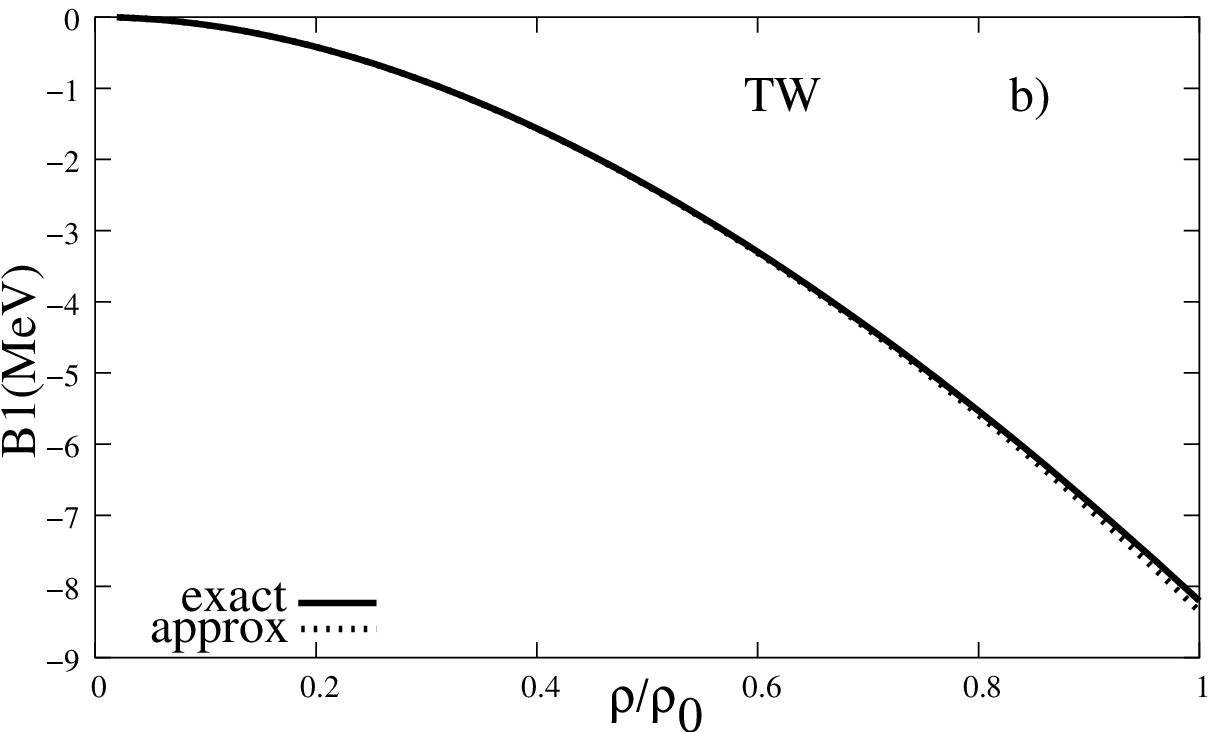} &
\includegraphics[width=5.5cm]{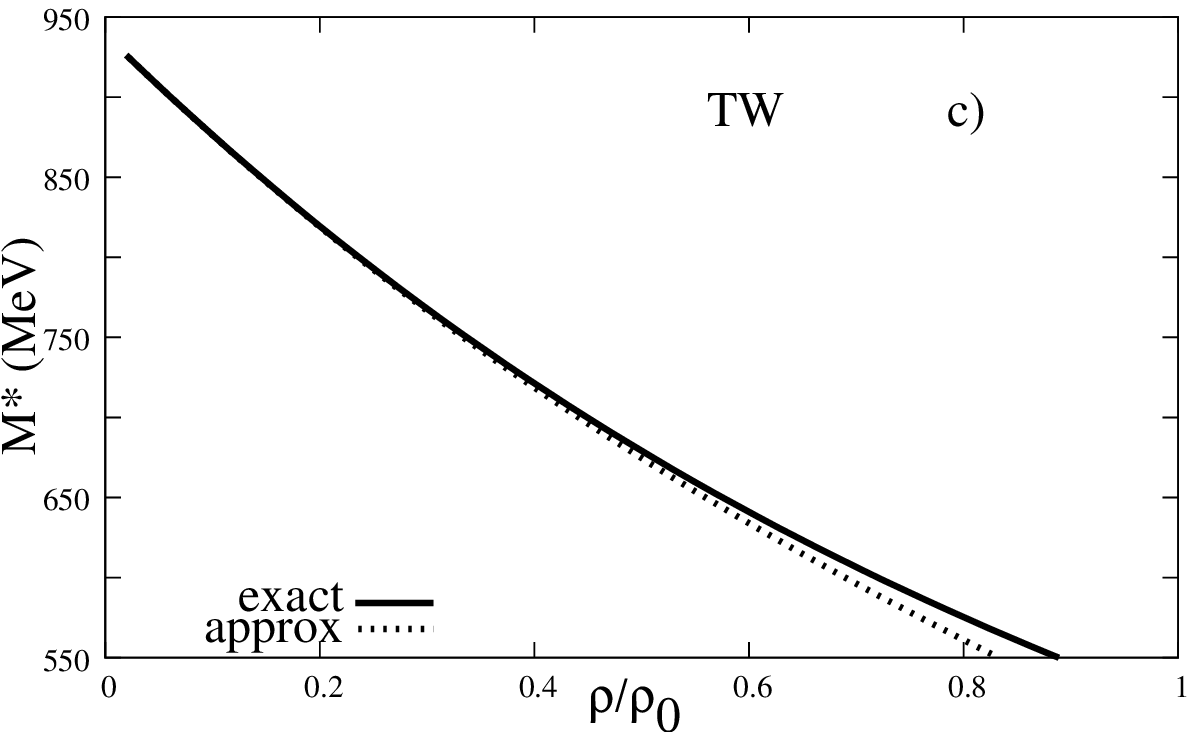}\\

\includegraphics[width=5.5cm]{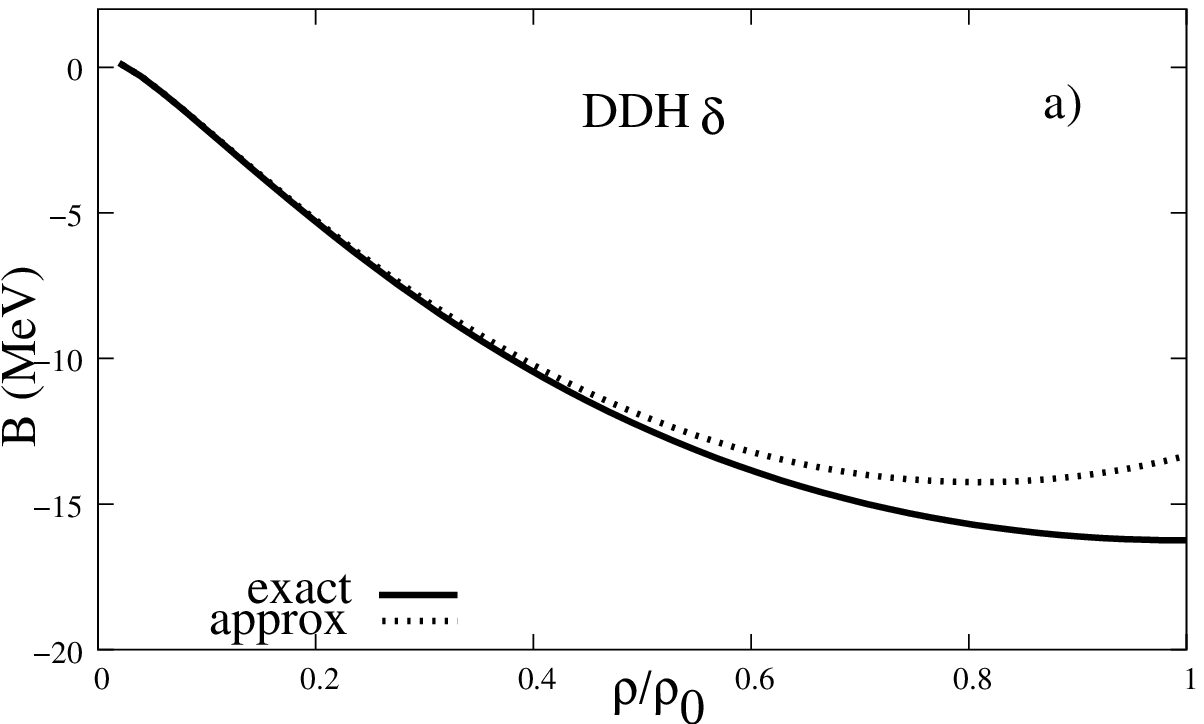} &
\includegraphics[width=5.5cm]{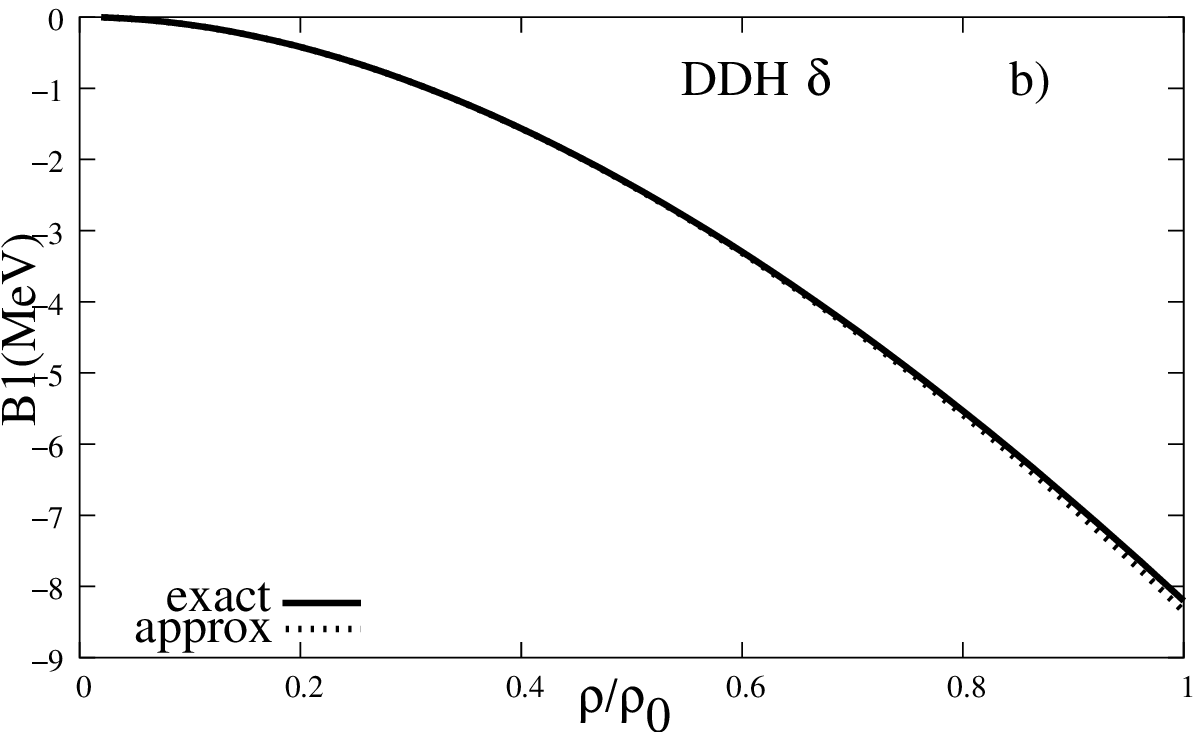} &
\includegraphics[width=5.5cm]{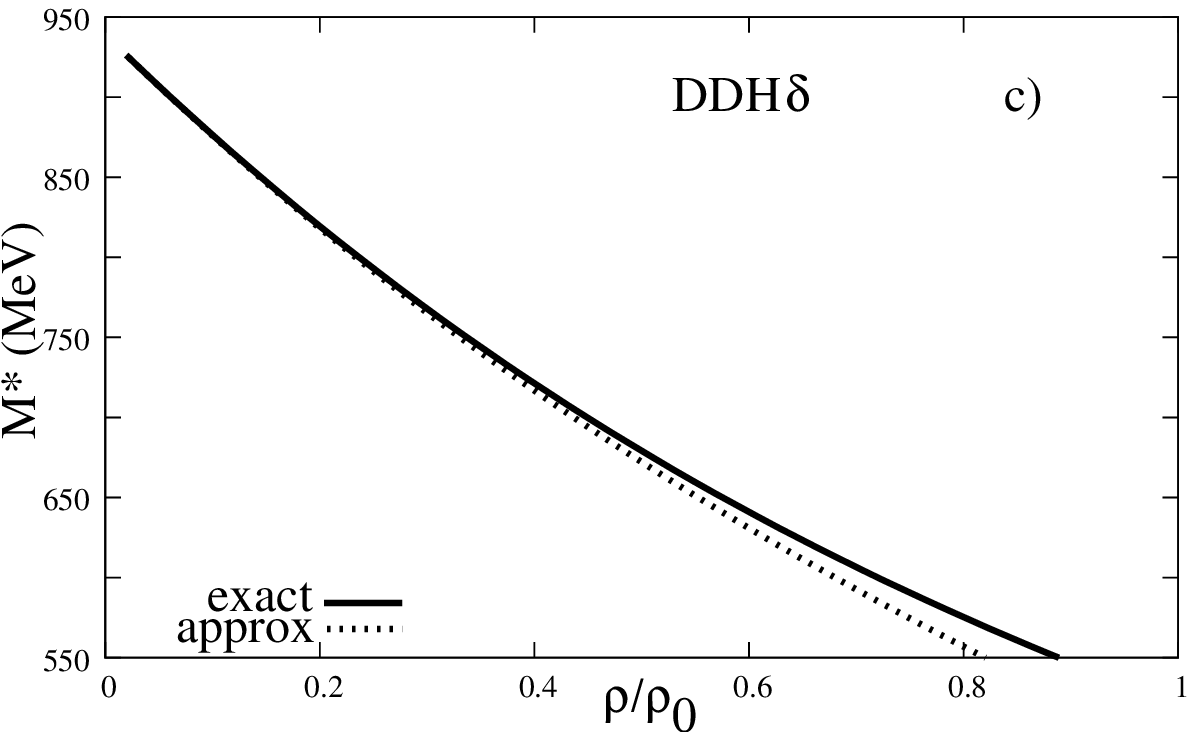}\\
\end{tabular}
\end{center}
\caption{Comparison between several exact and approximate physical quantities:
  a) binding energy density, b) $B_1(\rho)$ coefficient and
c) effective mass $M^*$ for relativistic models with density
dependent couplings.}\label{Bb1mdd}
\end{figure}

\begin{figure}[t]
\begin{center}
\begin{tabular}{cc}
\includegraphics[width=6.5cm,angle=0]{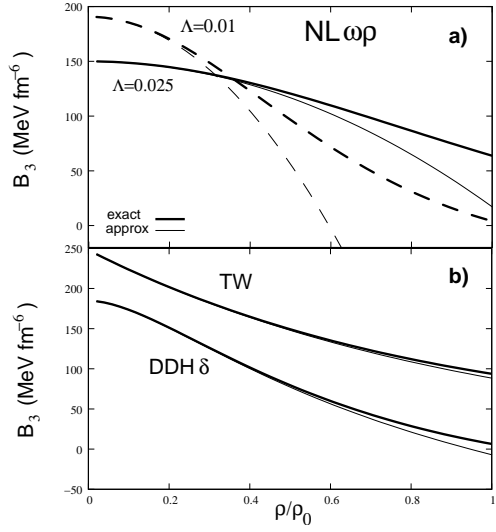} \\
\end{tabular}
\end{center}
\caption{$B_3(\rho)$ coefficient for a) a non-linear coupling
model (NL$\omega\rho$) and  b) two density dependent models  TW and
DDH$\delta$.}\label{B33rm}
\end{figure}

In Figs. \ref{Bb1mdd} we  show the binding energy, the
coefficient $B_1(\rho)$ and the effective mass obtained in the
non-relativistic approximation and the corresponding exact values.
We see that the low density non-relativistic approximation works
quite well both with and without the $\delta$ meson.
The same is true for the $B_3$ contribution.
In Fig.  \ref{B33rm}b) the exact and approximate coefficients
$B_3(\rho)$ are plotted for these density dependent models (TW and
DDH$\delta$).

\subsection{\protect\smallskip Comparison between relativistic and Skyrme
functionals}

We now compare the non-relativistic functional obtained from the
relativistic models described so far with the Skyrme functional.

In Table \ref{para0} we have collected the terms $B_1$,  $B_3$, $M_1$
and $M_3$ for the models we have considered in the present work. We notice 
that for the
Skyrme forces the inverse of the effective mass is parametrized
according to (\ref{masseffS}) while for the  relativistic models we
take a similar expression for the effective mass,
Eq. (\ref{nlmeff}). This fact explains the minus sign difference
between the two types of models for the  $M_1$
and $M_3$ columns. Except for the Walecka model, all models have for
the isoscalar interaction contribution a parabolic term plus a higher
order term on the density. This second contribution appears explicitly through
 $\rho^3$  and $\rho^4$ terms, or implicitly through the density
dependence of the coupling parameters for TW and DDH$\delta$. The
isovector interaction contribution has a much poorer parametrization
in the relativistic models: $B_3$ is generally constant because $D_3=0$, except for the
NL$\omega\rho$ model and again the TW and DDH$\delta$ models due to
the density dependence of the coupling parameters.

In Table \ref{para} we show the values of these coefficients for
the models that we have discussed. One can see that $C_0$ and $C_{\rm eff}$
are of the same order of the corresponding parameters of some  Skyrme models
shown in \cite{chabanat,douchin}. { The $C_3$ coefficients are normally twice
  as large for relativistic than for non-relativistic models and the $D_0$
  coefficients for relativistic models are half of the coefficients of the
  non-relativistic ones.} An immediate conclusion already
referred is the poor parametrization of the isovector channel in the
relativistic models: for most models both $D_3$ and $D_{\rm eff}$ are
zero. Some comments with respect to the isoscalar channel are also in
order: for the relativistic models the scalar  kinetic contribution,
defined by $C_{\rm eff}$ is
higher. This is due to the smaller effective mass within these models. 
The saturation is possible with an overall larger binding for
the isoscalar channel. This channel has an attractive term 
from the two-body force and a repulsive three-body ($n$-body)
contribution.
A larger binding may be obtained  with a stronger two-body attractive
potential ($C_0$) or a weaker three-body contribution ($C_3$ together
with the $\sigma$ exponent).

The  coefficients of the density dependent models discussed before   
could not be included in the above table since they are not fixed 
quantities. In Figs. \ref{coef} the coefficients for these models, defined in 
equations (\ref{b1tw}-\ref{mtw}), are plotted. Their values agree with
the ones already given in Table \ref{para} for density values 
$\rho/\rho_0>0.2$.

In order to better compare the relativistic models, and in
particular the parametrizations with density dependent coupling
parameters, with several Skyrme force models we have also fitted
the exact $B_1$ and $B_3$ by the (\ref{B1S}) and (\ref{B3S})
expressions in the density range 0 -- 0.1 fm$^{-3}$ using,
whenever appropriate, the same value for $\sigma$ in both
expressions. In Table \ref{para1} we give the results of these
fits.

Some conclusions can be drawn from the values in Tables \ref{para}
and \ref{para1}.  First of all we conclude that the isoscalar
channel in relativistic mean field models has a quite complicated
density dependence, both the interaction and momentum dependent term. In 
Table \ref{para} we only give the coefficients of the first terms of the 
expansion, which, as
 discussed before, works quite
 well at low subsaturation densities for QHD-II and  TM1 but not so well for
 NL3, NL$\delta$ and NL$\omega\rho$.  Using expressions
 (\ref{B1S}) and (\ref{B3S}) to parametrize the isoscalar and isovector 
interaction term we have obtained 
 for all the relativistic models a non-integer coefficient $\sigma$ smaller
 than 1 except for the 
 QHD-II. This is a special case which, as we know, does not describe correctly 
nuclear matter properties, namely, it predicts a
very large compressibility.  All other $\sigma$ values are smaller than 1 but 
not so small as the corresponding parameter in Skyrme forces with good performances which are
 generally below 0.2. This small value of $\sigma$ in Skyrme forces controls the
 compressibility and is generally taken equal to 1/6.
 Another term which has a very systematic behavior is the isoscalar momentum
 dependent term described by $C_{\rm eff}$: this coefficient, except for the 
NL$\delta$, is larger than
 50 and maybe as large as $\sim 100$ MeV fm$^5$ for the TW. This occurs at low
   densities and is compensated by an extra binding coming from de attractive
   term described by $C_0$.
Non-relativistic models have a similar behaviour, i.e., 
 $C_{\rm eff} \sim 50$ MeV fm$^5$. It should be noticed however that TW has a 
richer density dependence
 which is parametrized by the coefficient $C_{{\rm eff},3}$ not present in the
 Skyrme forces. On the other hand, relativistic models have generally a very 
simple isovector channel. The inclusion of the $\delta$-meson brings in the 
extra degree of freedom missing in the momentum dependent
 terms but not in the interaction term if only linear terms are included for
 the mesons.
%in the model. 
Models with density dependent coefficients such as the TW include automatically a larger density
 dependence in the isovector term of the interaction. When we compare $D_3$ for TW
 with $D_3$ for SLy230a, NRAPR or LNS we observe a similar behavior.

\section{Conclusions}

In the present work, we have 
compared relativistic mean-field nuclear models the relativistic models with
 the
non-relativistic Skyrme models. We have shown that for the isoscalar
channel the relativistic models behave in a similar way, and
generally different from the non-relativistic description.  This is true
for the binding energy, isoscalar interaction term and effective
mass. The
relativistic density dependent  models give the closer 
 description to the one obtained by the  non-relativistic models.   

The isovector channel has proved to be a different problem:
there is a quite big discrepancy even between models within the same
framework. This is the least known part of the nuclear interaction
which we expect to determine with the future radioactive beams.
Relativistic models have generally a very poor description of the this channel.

We have next tried to obtain a low density expansion of the
relativistic models with a parametrization similar to the one used for
the Skyrme interactions. The energy functional of the
relativistic models depends not only on the isoscalar and isovector
particle and kinetic densities but also on the isoscalar and isovector
 scalar densities. Only in the  low density regime these densities
reduce to the respective particle densites. We have shown that for
some models already in  the subsaturation
density expansions  it is necessary
to include many terms  in order to get good agreement. For
the low density range for which there is a good
agreement between the low density expansion and the exact values, we
have shown that some of the %parameters 
coefficients of the relativistic and the 
non-relativistic models are of the same order of magnitude. However
we have shown that for the first ones the scalar  kinetic contribution is
higher. The saturation is possible with an overall larger binding for
the isoscalar channel which is due to a stronger two-body attractive
potential or a weaker three-body contribution.

\section*{ACKNOWLEDGMENTS}

This work was partially supported by CNPq (Brazil),
CAPES(Brazil)/GRICES (Portugal) under project 100/03 and FEDER/FCT
(Portugal) under the projects POCTI/FP/63419/2005 and
POCTI/FP/63918/2005.

\begin{table}
\caption{Expressions for the coefficients $B_1(\rho)$,
$B_3(\rho)$, $M_1(\rho)$ and $M_3(\rho)$ . For the Skyrme
parametrization we take (\ref{masseffS}) which refers to ${M^*}^{-1}$.}  \label{para0}
\vspace{0.3cm}
\begin{center}
\begin{tabular}{lcccc}
\hline 
$\phantom{00}$model & $ \phantom{00}B_1(\rho)$ & $\phantom{00}B_3(\rho)$ &$\phantom{00}
M_1(\rho)/\rho$ & $\phantom{00}M_3(\rho)$ \\
\hline
Skyrme &$C_0 \rho^2+C_3 \rho^{\sigma+2}$ &$D_0+D_3
\rho^{\sigma}$ &$2C_{\rm eff}$ &$2D_{\rm eff}$ \\
\hline
QHD-II & $(c_v-c_s) \rho^2$& $c_{\rho}$ & $-2 c_s $ &  \\

NL3, TM1 & $(c_v-c_s) \rho^2+c_{s3} \rho^3 +(c_{s4}+c_{v4})\rho^4$
& $c_{\rho}$ & $-2 c_s $ \\
NL$\delta$ & $(c_v-c_s) \rho^2+c_{s3} \rho^3
+(c_{s4}+c_{v4})\rho^4$ & $c_{\rho}-c_{\delta}$ & $-2 c_s$&
$-2 c_{\delta}$\\

NL$\omega \rho$, NL$\sigma \omega$ & $(c_v-c_s) \rho^2+c_{s3}
\rho^3 +c_{s4}\rho^4$ &
$c_{\rho}-(3 c_{v \rho} +c_{s \rho}) \rho^2$ & $-2 c_s $ \\

TW & $(C_v-C_s) \rho^2$  &  $C_{\rho}$ & $-2 C_s $ \\

DDH$\delta$ & $(C_v-C_s) \rho^2$ & $C_{\rho}-C_{\delta}$ & $-2 C_s
$& $-2 C_{\delta}$\\
\hline
\end{tabular}
\end{center}
\end{table}

\vspace*{6cm}

\begin{table}
\caption{Coefficients $C_i$ and $D_i$ obtained in the present work
and results from several Skyrme models.} \label{para}
\vspace{0.3cm}
\begin{center}
\begin{tabular}{lccccccc}
\hline model & $C_0$ & $C_3$ & $D_0$ & $D_3$ & $C_{\rm eff}$ &
$D_{\rm eff}$ &
$\sigma$\\
& (MeV fm$^{3}$) & (MeV fm$^{3+3 \sigma}$) & (MeV fm$^{3}$) &
(MeV fm$^{3+3 \sigma}$) & (MeV fm$^{5}$) & (MeV fm$^{5}$) & \\
\hline
QHD-II & -308.83 &  0 & 59.30 & 0 & 50.97 &  0 & 1  \\
NL3 & -511.34 & 2482.69 & 131.44 & 0 & 68.01  & 0 & 1\\
NL$\omega\rho$,$\Lambda_v=0.01$ & -511.34 & 2482.69 & 149.22 & -5975.74 $(\sigma=2)$
& 68.01
& 0 & 1\\
NL$\omega\rho$,$\Lambda_v=0.025$ & -511.34 & 2482.69 & 189.76 & -24161.25 $(\sigma=2)$
& 68.01
& 0 & 1\\
TM1 & -479.21 & 1571.51 & 138.34 &  0 & 64.90 &  0  & 1\\
NL$\delta$ & -482.07 &  2369.38 &  63.82 &  0 & 44.64 &  10.81 & 1 \\
\hline \hline
SIII \cite{chabanat}    & -426.28 & 875 & 268.08 & 0 & 44.38 & -30.63 & 1\\
Sk1$^{\prime}$ \cite{douchin} & -396.49 & 903.97 & 208.42 & 0 &
12.98 & -20.99
& 1\\
SLy230a \cite{chabanat} & -933.84 & 862.69 & 1015.89 & -1392.89 &
56.37 &
56.37 & 1/6  \\
NRAPR \cite{sple} & -1019.89 & 940.13 & 449.80 & -398.68 & 57.015
& -27.992 &
0.14416 \\
LNS \cite{lns} & -931.86 & 911.76 & 349.62 & -283.17 & 25.05 &
-19.5 & 1/6 \\ \hline
\end{tabular}
\end{center}
\end{table}

%\newpage
%.
%\newpage

\begin{table}
\caption{Coefficients obtained from the fitting to the exact $B_1$
and $B_3$ expressions} \label{para1} \vspace{0.3cm}
\begin{center}
\begin{tabular}{lcccccccc}
\hline model & $C_0$ & $C_3$ & $D_0$ & $D_3$ & $C_{\rm eff,0}$
&$C_{\rm
  eff,3}$ & $D_{\rm eff}$ &
$\sigma$\\
& (MeV fm$^{3}$) & (MeV fm$^{3+3 \sigma}$) & (MeV fm$^{3}$) & (MeV
fm$^{3+3 \sigma}$) & (MeV fm$^{5}$)&(MeV fm$^{3+3 \sigma}$) &
(MeV fm$^{5}$) & \\
\hline
QHD-II & -297.0 &  253.625 & 59.30 & 0 & 50.97 &  0 &0& 2.5 \\
NL3 & -625.25 & 546.28 & 131.44 & 0 & 68.01  & 0 & 0&0.318\\
TM1 & -482.84 & 546.07 & 138.34 &  0 & 64.90 &  0  & 0&0.606\\
NL$\delta$ & -594.83 &  802.09 &  63.82 &  0 & 44.64 & 0& 10.81 & 0.536 \\
TW & -486.09 &  613.50 &  251.92 &  -746.72 & 96.29  &-136.80 &0 &
0.767 \\
\hline
\end{tabular}
\end{center}
\end{table}

\newpage
.
\newpage

\newpage

\end{document}